\documentclass[pdflatex, 11pt]{article}

\usepackage{mathrsfs}
\usepackage{amssymb}
\usepackage{amsmath}
\usepackage[dvipdfmx]{graphicx}
\usepackage{natbib}
\usepackage{setspace}
\usepackage{float}
\usepackage{comment}
\usepackage[title]{appendix}
\usepackage{booktabs}
\usepackage{color}
\usepackage{times}
\usepackage{url}
\usepackage{algorithm}
\usepackage{algpseudocode}
\usepackage{geometry}

\newcommand{\va}{\mathbf{a}}
\newcommand{\vb}{\mathbf{b}}

\newcommand{\ve}{\mathbf{e}}

\newcommand{\vm}{\mathbf{m}}

\newcommand{\vu}{\mathbf{u}}

\newcommand{\vx}{\mathbf{x}}

\newcommand{\vz}{\mathbf{z}}

\renewcommand{\d}{\text{d}}

\newcommand{\mB}{\mathbf{B}}

\newcommand{\mD}{\mathbf{D}}

\newcommand{\mI}{\mathbf{I}}

\newcommand{\mM}{\mathbf{M}}
\newcommand{\mP}{\mathbf{P}}

\newcommand{\mS}{\mathbf{S}}

\newcommand{\vbeta}{\text{\boldmath{$\beta$}}}

\newcommand{\veta}{\text{\boldmath{$\eta$}}}
\newcommand{\vgamma}{\text{\boldmath{$\gamma$}}}

\newcommand{\vmu}{\text{\boldmath{$\mu$}}}

\newcommand{\vtheta}{\text{\boldmath{$\theta$}}}
\newcommand{\vvartheta}{\text{\boldmath{$\vartheta$}}}

\newcommand{\mTheta}{\mathbf{\Theta}}
\newcommand{\mSigma}{\mathbf{\Sigma}}

\newcommand{\floor}[1]{\lfloor #1 \rfloor}

\newcommand{\los}{\texttt{LoS}}
\newcommand{\age}{\texttt{age}}
\newcommand{\hr}{\texttt{hr}}
\newcommand{\bmi}{\texttt{bmi}}
\newcommand{\sysbp}{\texttt{sysbp}}
\newcommand{\diasbp}{\texttt{diasbp}}

\usepackage[figuresright]{rotating}

\title{\textbf{Flexible Bayesian Quantile Regression for Counts \\ via Generative Modeling}}
\date{}
\author{
}

\begin{document}
\maketitle
\doublespacing

\vspace{-1.5cm}
\begin{center}
Yuta Yamauchi$^{1}$, 
Genya Kobayashi$^{2}$\footnote{Author of correspondance: \texttt{gkobayashi@meiji.ac.jp}} and 
Shonosuke Sugasawa$^{3}$
\end{center}

\noindent
$^1$Department of Economics, Nagoya University\\
$^2$School of Commerce, Meiji University\\
$^3$Department of Economics, Keio University

\begin{abstract}
Count data frequently arises in biomedical applications, such as the length of hospital stay. 
However, their discrete nature poses significant challenges for appropriately modeling conditional quantiles, which are crucial for understanding heterogeneous effects and variability in outcomes. 
To solve the practical difficulty, we propose a novel general Bayesian framework for quantile regression tailored to count data.
We seek the regression parameter on the conditional quantile by minimizing the expected loss with respect to the distribution of the conditional quantile of the latent continuous variable associated with the observed count response variable. 
By modeling the unknown conditional distribution through a Bayesian nonparametric kernel mixture for the joint distribution of the count response and covariates, we obtain the posterior distribution of the regression parameter via a simple optimization.
We numerically demonstrate that the proposed method improves bias and estimation accuracy of the existing crude approaches to count quantile regression.
Furthermore, we analyze the length of hospital stay for acute myocardial infarction and demonstrate that the proposed method gives more interpretable and flexible results than the existing ones. 

\end{abstract}


\newpage
\section{Introduction}

Quantile regression has emerged as a powerful tool for understanding the distribution of a response variable conditioned on explanatory variables. 
Unlike ordinary linear regression, which focuses on the mean of the conditional distribution, quantile regression provides insights into various quantiles of the distribution, offering a more comprehensive view of the relationship between variables \citep{koenker2005quantile,qr_handbook}. 
This approach is particularly advantageous when the distribution of the response variable is skewed or when heteroscedasticity exists as a function of explanatory variables, which frequently appear in biomedical applications \citep[e.g.,][]{faddy2009modeling}.

Traditional quantile regression methods primarily deal with continuous response variables. 
Although many biological and medical applications involve count data such as frequencies of genetic markers from sequencing data \citep{Lee_bio}, prevalences of diseases \citep{Koba, Arima}, 
lengths of hospital stays \citep{Wang}, and numbers of prescribed drugs \citep{geraci2022mid}, quantile regression for count data still presents several unsolved challenges. 
Traditional methods, such as jittering and maximum score estimation, have laid the groundwork but also highlighted several limitations \citep{Manski75, Manski85, Benoit, Machado2005, LeeTereza, frumento2021parametric, padellini2018model, lamarche2021conditional, liu_yu, geraci2022mid}. 
Jittering, for example, smoothes discrete data by adding noise.
However, this can significantly affect small values and can suffer from the usual problem of quantile crossing. 
Also, \cite{Alahmadi} argues that crossing is not solved by simply discretizing the asymmetric Laplace likelihood as considered by \cite{liu_yu}. 
\cite{ilienko2013continuous} introduced a continuous Poisson distribution with the cumulative distribution expressed using the floor function. 
\cite{lamarche2021conditional} proposed a three-step approach that first fits a regression model with a usual discrete distribution, calculates (imputes) conditional quantiles on a continuous scale, and then regresses the estimated quantile value on a set of covariates. 
A similar approach was also considered by \cite{padellini2018model}. 
\cite{Alahmadi} considered further model extension within the parametric framework. 
Their parametric continuous-discrete approach faces the usual limitations of the parametric model in both distributional assumptions and specification of regression functions, which may overlook heterogeneity in covariate effects on different quantiles, as shown in our application. 

To overcome the aforementioned issues, we propose a novel approach to Bayesian quantile regression for count data.
The essential part is to define the covariate effect on the conditional quantile as the minimizer of the expected loss function, where the expectation is taken with respect to the joint empirical distribution of the covariates and conditional quantile of the continuous counterpart to the count response.
We then consider posterior inference on the defined parameter in the general Bayesian framework \citep{bissiri2016general}. 
We propose a Bayesian nonparametric kernel mixture of truncated rounded Gaussians for the joint distribution of the count response and covariate by extending the methods developed by \cite{canale2011bayesian} and \cite{canale2017robustifying}. 
Such generative modeling allows us to fully capture the underlying structure between the covariates and count response, unlike the existing parametric approaches of \cite{padellini2018model} and \cite{lamarche2021conditional}.  
This leads to a flexible estimation of conditional quantiles that can successfully capture heterogeneity across different quantile levels. 
In fact, our application to the length of hospital stay demonstrates that the proposed method can detect the heterogeneity of covariate effects across different quantiles, which is missed by the existing method. 
We also demonstrate the efficacy of the proposed method through numerical experiments. 
The results showcase its superiority in accurately recovering the underlying quantiles. 
Hence, the proposed method enhances the accuracy of quantile estimates and improves the interpretability of the relationship between the response and explanatory variables, making it a valuable tool for researchers and practitioners working with count data in various medical applications.

The rest of the paper is organized as follows. 
Section~\ref{sec:model} introduces the general Bayesian approach to count quantile regression through generative modeling based on the joint distribution of response and explanatory variables. 
In Sections~\ref{sec:sim} and \ref{sec:empirical}, we demonstrate the efficacy of the proposed method through simulation and real data analysis. 
Finally, concluding remarks are given in Section~\ref{sec:conc}.

\section{Count quantile regression via generative modeling}\label{sec:model}

\subsection{General Bayesian approach to inference on quantile regression function}\label{sec:generalbayes}
Let us first denote the observed count response variable by $y$. 
We aim to quantify the effect of the $p$-dimensional vector of covariates $\vx$ on the various parts of the distribution of $y$ via regression and propose a general Bayesian approach. 

Suppose that the observable response variable $y$ takes nonnegative integer values: $y \in \{0, 1, \ldots\}$. 
Let $y_{\tau}^*$ denote $\tau$th conditional quantile of a continuous variable $y^*$ given the $p$-dimensional vector of covariates $\vx=(x_1,\dots,x_p)$. 
This latent continuous variable is associated with the observed count $y$ through
\begin{equation}\label{eqn:group}
y = g  \ \iff  \ y^* \in (g - 1, g], \quad g \in \{0, 1, \ldots\}.
\end{equation}

In this paper, rather than directly modeling a regression of the observed count $y$ on $\vx$, we propose to regress the latent continuous conditional quantile $y_{\tau}^*$ on $\vx$, similarly to \cite{padellini2018model} and \cite{lamarche2021conditional}.  
Let us denote the parameter of the regression model for $\tau$th continuous conditional quantile by $\vbeta_\tau$, which is defined as the minimizer of the expected loss given by 
\begin{equation}\label{eqn:parameter}
\vbeta_{\tau}\equiv \underset{\vbeta}{\operatorname{argmin}} \int L(\vbeta; y_{\tau}^{*}, \vx) F(y_{\tau}^{*},\vx) \d y_{\tau}^{*}\d \vx,
\end{equation}
where $F(y_{\tau}^{*},\vx)$ is the true joint distribution of $y_{\tau}^{*}$ and $\vx$, and $L(\vbeta; y_{\tau}^{*}, \vx)$ is a loss function. 
The choice of $L(\vbeta; y_{\tau}^{*}, \vx)$ is arbitrary, and any penalization or regularization term can be included as in the usual regression analysis, depending on the purpose of the analysis and the characteristics of the data. 
For example, it is possible to include the LASSO penalty to introduce sparsity. 

In this paper, we consider the additive regression model to account for the nonlinear effect of the covariates on the continuous conditional quantile, and adopt the penalized squared loss function given by 
\begin{equation}\label{eqn:loss}
L(\vbeta; y_{\tau}^{*}, \vx)
=\bigg\{ y_{\tau}^{*} - \sum_{j=1}^P \mB(x_{j})' \vbeta_{j} \bigg\}^2 + \sum_{j=1}^P\lambda_{\tau,j} \vbeta_j' \mP_{\tau,j} \vbeta_j,
\end{equation}
\label{page:D}
where $\mB(\cdot)$ is the basis function, $\vbeta=(\vbeta_1',\dots,\vbeta_P')'$ is the collection of the basis coefficients, $\mP_{\tau}$ is the penalty matrix, and $\lambda_\tau$ is the parameter to control the smoothness of the nonlinear effect. 
In this paper, $\mB(\cdot)$ is the cubic B-spline basis function with the ten equi-distant knots and $\mP_{\tau,j}=\mD_{j}'\mD_j$ where $\mD_j$ being the matrix representation of forming the second order differences of the adjacent coefficients: $(\beta_{kj}-\beta_{k-1,j})-(\beta_{k-1,j}-\beta_{k-2,j})$. 
To compute $\vbeta_\tau$, we utilize the R package \texttt{SOP}.  
The package \texttt{SOP} uses the equivalent linear mixed model representation, and estimates the smoothness parameter $\lambda_\tau$ from the estimates of the variance parameters \citep[see also][]{sap, sop2}.

If the true distribution $F(y_{\tau}^{*},\vx)$ were known, one can exactly compute $\vbeta_{\tau}$ without uncertainty. 
However, since the true distribution is generally unknown, $F(y_{\tau}^{*},\vx)$ is subject to inference. 
This paper considers the form of inference given by $F(y_{\tau}^{*},\vx)=F(y_{\tau}^{*}|\vx)F_N(\vx)$ where $F_N(\vx)=N^{-1}\sum_{i=1}^N \delta_{\vx_i}$ is the empirical distribution of $\vx$ and $\delta_{\vx_i}$ is a Dirac measure on $\vx_i$. 
The conditional distribution $F(y_{\tau}^{*}|\vx)$ is a key target of inference, as the resulting regression estimates hinge on the flexibility in the model for $F(y_{\tau}^{*}|\vx)$. 
In the present paper, this conditional distribution is implied from the joint distribution of $y$ and $\vx$, for which a Bayesian nonparametric mixture is considered. 
From the standard Bayesian updating, the posterior predictive distribution of $F(y_\tau^*|\vx)$ is obtained. 
Therefore, the uncertainty in the inference for $F(y_{\tau}^{*}|\vx)$ is naturally reflected in the inference for $\vbeta_{\tau}$ in \eqref{eqn:parameter}.  
Such a construction gives a generalized version of the posterior distribution of $\vbeta_{\tau}$, known as 
general Bayesian updating \citep{bissiri2016general} or nonparametric Bayesian learning \citep{lyddon2018nonparametric}. 

Given the pairs of the observed count response and covariate, $(y_i,\vx_i)$ for $i=1,\dots,n$, where $\vx_i=(x_{1i},\dots,x_{pi})$, Algorithm~\ref{tab:algo1} produces a random sample from the general posterior of $\vbeta_\tau$ in \eqref{eqn:parameter}.   
Step~1 of the algorithm fits a Bayesian nonparametric mixture model using a Markov chain Monte Carlo (MCMC) algorithm. 
For each MCMC draw of the joint distribution of $(y_i,\vx_i)$, in Step~2 of the algorithm, we can compute a posterior draw of $\tau$th latent continuous conditional quantile, denoted by $y^{*(s)}_{\tau,i}$ with the parenthesized superscript. 
In Step~3, $y^{*(s)}_{\tau,i}$ is regressed on $\vx_i$ to compute a posterior draw of the regression parameter $\vbeta_\tau^{(s)}$, from which we obtain the regression function given $\vx$, denoted by $Q^{(s)}_{y^*}(\tau;\vx)=\sum_{j=1}^P \mB(x_j)' \vbeta_{\tau, j}^{(s)}$ for $s=1,\dots,S$. 
Using the posterior sample, we can compute the posterior means and credible intervals for quantities of interest. 
For example, a posterior draw of the conditional quantile for the count response is given by $Q^{(s)}_{y}(\tau;\vx)=\lceil Q^{(s)}_{y^*}(\tau;\vx)\rceil$  where $\lceil y\rceil:=\min\left\{z\in\mathbb{Z}: z\geq y\right\}$ is the ceiling function. 
We will explain the details of the steps in the subsequent sections.

\begin{algorithm}
\begin{algorithmic}[1]
\Require {Count response $y_i$ and covariates $\vx_i$ for $i=1,\dots,N$.  }

\item
Fit the nonparametric mixture model (Section~\ref{sec:joint_model}) 
\begin{equation*}
\begin{split}
    &y_i = g \quad \text{iff}\quad  y_i^*\in(g-1,g]\\
    &\vx_i|y_i^*,\vmu_{\vx,i},\mSigma_{c,i},\veta_{i}\sim N(\vmu_{\vx,i}+\sigma_{y,i}^{-2}\veta_i(y^*_i-\mu_{y,i}),\mSigma_{c,i})\\
    &y_i^*|\mu_{y,i},\sigma^2_{y,i} \sim TN_{1}(\mu_{y,i}, \sigma^2_{y,i},-1)\\
    &\vtheta_i|G \sim G,\quad 
    G|\phi, \vartheta, G_0 \sim PY(\phi,\vartheta;G_0),
\end{split}
\end{equation*}
using the MCMC algorithm described in Section~\ref{sec:mcmc} of the supplementary material. 

\item 
Generate a random sample $y_{\tau,i}^{* (s)}$ from the posterior predictive distribution of $F(y_{\tau,i}^{*}|\vx_i)$ for $i=1,\ldots,N$ and $s=1,\ldots,S$ by solving
\[
\frac{\int_{-1}^{y^{*(s)}_{\tau,i}} \int k(\vz_i;\vtheta)\d G^{(s)}(\vtheta)\d y^*}{\int k_\vx(\vx_i;\vtheta_\vx)\d G^{(s)}(\vtheta_\vx)}=\tau, 
\]
for $y_{\tau,i}^{* (s)}$ where $\vz_i=(y_i^*,\vx_i)$, using the method described in Section~\ref{sec:cq} of the supplementary material. 

\item 
For $s=1,\ldots,S$, solve the following optimization problem for $\vbeta_{\tau}^{(s)}$ using the package \texttt{SOP}:
\begin{equation}
\vbeta_{\tau}^{(s)}=\underset{\vbeta}{\operatorname{argmin}}\left[
\sum_{i=1}^n\bigg\{ y_{\tau,i}^{* (s)} - \sum_{j=1}^P \mB(x_{ij})' \vbeta_{j} \bigg\}^2 + \sum_{j=1}^P\lambda_{\tau,j} \vbeta_j' \mP_{\tau,j} \vbeta_j
\right].
\end{equation}

\Ensure Posterior draws of the coefficient parameter $\vbeta_{\tau}^{(s)}$ for $s=1,\dots,S$. 

\end{algorithmic}
\caption{General Bayesian framework to count quantile regression}
\label{tab:algo1}
\end{algorithm}

\subsection{Nonparametric mixture for the joint distribution $F(y^*,\vx)$}\label{sec:joint_model}
As in the previous section, the proposed general Bayesian approach requires the posterior predictive distribution of the latent conditional quantile $y_\tau^*$. 
Also, $F(y^*_\tau|\vx)$ needs to be sufficiently flexible to capture the underlying structure through the regression, and a parametric model such as \cite{padellini2018model} and \cite{lamarche2021conditional} would fail to do so. 
To this end, we nonparametrically model the joint distribution of the observed count response $y$ and covariate $\vx$, then obtain the posterior draws of the implied $F(y^*_\tau|\vx)$.

To obtain a flexible conditional distribution $F(y^*|\vx)$, we jointly model the distribution of $y^*$ and $\vx$ through the nonparametric mixture following \cite{canale2011bayesian} and \cite{canale2017robustifying}. 
Denoting the $P+1$-dimensional vector by $\vz = (y^*, \vx')'$, the joint density is given by
\begin{equation}\label{eqn:kernelmix}
f(\vz; G) = \int k(\vz;\vtheta) \d G(\vtheta), 
\end{equation}
where $k(\cdot ; \vtheta)$ denotes a parametric kernel density, and $G$ denotes a random mixing distribution.
Given the implied conditional density  $f(y^*|\vx)$, the conditional probability mass function for $y$ is given by
$
p(y=g|\vx)=\int_{g-1}^g f(y^*|\vx)dy^* 
$. 
Assuming the covariates $\vx \in \mathbb{R}^P$ are continuous for simplicity, we adopt the multivariate truncated normal kernel for $k(\vz;\vtheta)$: 
\begin{equation}\label{eqn:PYmix1}
\begin{split}
    &\vz|\vmu,\mSigma \sim TN_{P+1}(\vmu, \mSigma,\va),\\
    &\vmu,\mSigma|G \sim G,
\end{split}
\end{equation}
where $TN_{D} (\vmu,\mSigma,\vu)$ denotes the $D$-dimensional normal distribution with mean $\vmu$ and covariance matrix $\mSigma$ truncated on $(u_1,\infty)\times\cdots\times(u_D,\infty)$ with $\vu=(u_1,\dots,u_D)'$, the $(P+1)\times 1 $ vector $\va$ is given by $\va=(-1,-\infty,\dots,-\infty)'$, implying $y^*>-1$. 
Following \cite{canale2017robustifying}, as a flexible choice for the nonparametric prior for $G$, we choose the Pitman-Yor (PY) process \citep{perman1992size,pitman1997two,ishwaran2003generalized}. 
The PY process includes the Dirichlet process (DP) as a special case and is a flexible tool for density estimation and, especially, for clustering with robust inference on the number of components.

The proposed model extends the nonparametric modeling for count data proposed by \cite{canale2011bayesian} and \cite{canale2017robustifying} to modeling a joint distribution. 
\cite{canale2017robustifying} proposed the nonparametric mixture modeling using the normal density for the latent continuous variable as a kernel of the PY mixture. 
The resulting kernel for the discrete response was referred to as the rounded Gaussian (RG) kernel. 
They demonstrated the rigidity of the Poisson kernel and the flexibility of the RG kernel and PY mixture. 

Furthermore, in our model, the kernel for $\vz$ is truncated such that $y^*>-1$. 
While a natural extension of the RG kernel for count data to the joint kernel for $\vz$ would have adopted the unrestricted multivariate normal kernel, our multivariate truncated normal kernel is an important choice in the present context of quantile estimation for count data because we are interested in the conditional quantile of the latent continuous variable $y_{\tau}^*$. 
Under the RG kernel of \cite{canale2017robustifying}, the support of $y^*$ associated with the observation $y=0$ is given by $y^*\leq 0$, as opposed to $(-1,0]$ in our model, and thus $y_{\tau}^*$ for a lower $\tau$ can take any nonpositive value, which causes numerical instability. 
Therefore, using the truncated kernel maintains the interpretability of $y_{\tau}^*$ by avoiding numerical instability.

A useful conditional conjugacy for the component parameters, $\vmu_k$ and $\mSigma_k$, of the multivariate truncated normal kernel is not available. 
Therefore, directly working on the formulation in \eqref{eqn:PYmix1} leads to full conditional distributions of $\mu$ and $\mSigma$, which are not in standard forms. 
Sampling these parameters using, for example, the Metropolis-Hastings (MH) algorithm typically requires tuning and does not work well for a nonparametric mixture model. \label{page:repara}
To facilitate the posterior computation, we consider an alternative but equivalent representation. 
To this end, the mean vector $\vmu$ and covariance matrix $\mSigma$ are partitioned as
$\vmu=(\mu_{y},\vmu_{\vx}')'$, 
$
\mSigma=
\begin{pmatrix}
    \sigma_{y}^2&\veta'\\
    \veta_i&\mSigma_{\vx}
\end{pmatrix}
$, 
and \eqref{eqn:kernelmix} is written as
\begin{equation}\label{eqn:kernelmix2}
f(\vz;G)=\int k_{y^*}(y^*;\mu_{y},\sigma_{y}^2)k_{\vx|y^*}(\vx|y^*;\vmu_{\vx},\mSigma_{c},\veta)\d G(\vtheta),
\end{equation}
where $\vtheta=(\mu_{y},\sigma_{y}^2,\vmu_{\vx},\mSigma_{c},\veta)$
, $k_{y^*}(y^*;\mu_{y},\sigma_{y}^2)$ is the marginal kernel for $y^*$ which is $TN_1(\mu_{y},\sigma_{y}^2,-1)$, $k_{\vx|y^*}(\vx|y^*;\vmu_{\vx},\mSigma_{c},\veta)$ is the conditional kernel for $\vx$ given $y^*$ which is the $P$-dimensional normal $N(\vmu_{\vx}+\sigma_{y}^{-2}\veta(y^*-\mu_{y}),\mSigma_{c})$. 
Note that $\mSigma_{\vx}$ can be recovered from $\mSigma_{\vx}=\mSigma_{c}+\sigma_{y}^{-2}\veta\veta'$. 
Then, given the data $(y_i,\vx_i)$ for $i=1,\dots,n$, we work on the following hierarchical model: 
\begin{equation}\label{eqn:PYmix2}
\begin{split}
    &y_i = g \quad \text{iff}\quad  y_i^*\in(g-1,g]\\
    &\vx_i|y_i^*,\vmu_{\vx,i},\mSigma_{c,i},\veta_{i}\sim N(\vmu_{\vx,i}+\sigma_{y,i}^{-2}\veta_i(y^*_i-\mu_{y,i}),\mSigma_{c,i})\\
    &y_i^*|\mu_{y,i},\sigma^2_{y,i} \sim TN_{1}(\mu_{y,i}, \sigma^2_{y,i},-1)\\
    &\vtheta_i|G \sim G,\quad
    G|\phi, \vartheta, G_0 \sim PY(\phi,\vartheta;G_0)
\end{split}
\end{equation}
where $\vtheta_i=(\mu_{y,i},\sigma_{y,i}^2,\vmu_{\vx,i},\mSigma_{c,i},\veta_i)$, $PY(\phi, \vartheta; G_0)$ denotes the PY process with the discount parameter $\phi \in [0, 1]$, strength parameter $\vartheta > -\phi$, and base measure $G_0$. 
The base measure is the product of the normal, inverse gamma, and inverse Wishart distributions:
$\mu_y\sim N(\mu_{y0}, \sigma^2_{y0})$,
$\sigma_y^2\sim IG(k_0, t_0)$,
$\vmu_\vx\sim N(\vmu_{\vx0},\mSigma_{\vx0})$,
$\mSigma_c\sim IW(n_0, \mS_0)$,
$\veta\sim N(\vmu_{\eta 0}, \mSigma_{\eta 0})$, 
where $IG(k_0, t_0)$ denotes the inverse gamma distribution with the shape parameter $k_0$ and scale parameter $t_0$, and $IW(n_0,\mS_0)$ denotes the inverse Wishart distribution with $n_0$ degrees of freedom and scale matrix $\mS_0$. 

The model is estimated using the MCMC algorithm within the Bayesian framework. 
We employ the importance conditional sampling algorithm of \cite{canale22} for clustering the observations based on the PY process, and cluster-specific parameters are updated using the kernel representation \eqref{eqn:kernelmix2} and \eqref{eqn:PYmix2}. 
The detail of the MCMC algorithm is provided in Appendix~\ref{sec:mcmc} of the supplementary material.

\subsection{Continuous conditional quantiles}\label{sec:quantile}
Based on the joint modeling method for $\vz$ described in Section~\ref{sec:joint_model}, we obtain the posterior samples of the continuous conditional quantiles, $y_{\tau}^*$, as in \cite{taddy2010bayesian}. 
Let $k_\vx(\cdot; \vtheta_\vx)$ denote the marginal kernel of $\vx$ with the associated parameter $\vtheta_\vx=(\vmu_\vx,\mSigma_\vx)$, which is $N(\vmu_\vx, \mSigma_\vx)$. 
The conditional density of \( y^* \)  and the conditional distribution function are, respectively, given by
\begin{equation}\label{eqn:cond}
f(y^*|\vx;G) = \frac{\int k(\vz; \vtheta) \d G(\vtheta)}{\int k_\vx(\vx; \vtheta_\vx) \d G(\vtheta_\vx)}, \quad 
F(y_{\tau}^* | \vx;G) = \int_{-1}^{y^*_{\tau}} f(y^*|\vx;G) \d y^*. 
\end{equation}
Thus, the $\tau$th conditional quantile $y^*_\tau$ satisfies 
$
F(y_{\tau}^* | \vx;G) 
= \tau. 
$
For $s$th posterior draw of $G$, denoted by $G^{(s)}$, $F(\cdot | \vx_i;G^{(s)})$ is obtained for $i=1,\dots,n$. 
By numerically solving $F(y_{\tau,i}^* | \vx_i;G^{(s)})= \tau$ given the covariate value $\vx_i$, we generate the posterior draw of $y^{*(s)}_{\tau,i}$ for $s=1,\dots,S$. 
See Appendix~\ref{sec:cq} of the supplementary material for the computational details.

\subsection{Connection to the existing approach}\label{sec:parametric}
We summarize the advantages of the proposed method in comparison to the existing parametric approach of  
\cite{padellini2018model,lamarche2021conditional}. 
To this end, we also describe their approach in detail. 

They used a continuous approximation to distribution functions
for count data within a class of parametric models as an alternative smoothing method to the jittering approach.
For example, suppose the count variable $Y$ follows the Poisson distribution with the distribution function denoted by $F_Y(\cdot)$:
\begin{eqnarray}
    Y\sim {Poi}(\lambda),\quad F_Y(y) = \frac{\Gamma(\floor{y} + 1, \lambda)}{\Gamma(\floor{y} + 1)}, \quad y\geq 0, \nonumber
\end{eqnarray}
where the floor function $\floor{y} := \max\{z\in \mathbb{Z}:z\leq y\}$, $\Gamma(z)$ is the gamma function, and $\Gamma(z,a)$ is the incomplete gamma function $\int_a^\infty e^{-z}t^{z-1}dz$.
Then, the distribution function of the continuous Poisson distribution, denoted by $cPoi(\lambda)$, for the interpolated count, $Y^*$, can be derived as follows \citep{ilienko2013continuous}:
\begin{equation}
Y^* \sim {cPoi}(\lambda), \quad F_{Y^*}(y^*) =  \frac{\Gamma(y^* + 1, \lambda)}{\Gamma(y^* + 1)},\quad y^* > -1\nonumber. 
\end{equation}
Figure~\ref{fig:poi} of the supplementary material shows the distribution functions of the discrete Poisson and continuous Poisson distributions. 

In the case of the Poisson regression model $y_i|\vx_i\sim Poi(\exp(\vx_i'\vtheta)),\ i=1,\dots,n$, for example, the first step estimates the usual Poisson regression model to obtain the estimate $\hat{\vtheta}$. 
Given a user-specified $\tau\in(0,1)$,  based on the continuous version $cPoi(\exp(\vx_i'\hat{\vtheta}))$, the second step solves for $y_{\tau,i}^*$
\begin{equation}
\frac{\Gamma(y_{\tau,i}^*+1,\exp(\vx_i'\hat{\vtheta}))}{\Gamma(y_{\tau,i}^*+1)}=\tau,\quad i=1,\dots,n.
\end{equation}
Finally, the third step regresses $y_{\tau}^*$ on $\vx$ to obtain the effect on $\tau$th quantile by minimizing a loss function such as \eqref{eqn:loss}. 

Algorithm~\ref{tab:algo2} of the supplementary material summarizes this interpolation framework to estimate the covariate effect on the quantile of the continuous version of the discrete distribution proposed by \cite{lamarche2021conditional}.

This method, based on the continuous counterpart, is a natural alternative to jittering for smoothing discrete responses. 
However, this approach requires an assumption about the distribution and the regression function. 
For the distributional assumption, one can use the continuous Poisson and negative binomial distributions introduced in \cite{padellini2018model}. 
However, these continuous counterparts have the same problem of inflexibility as the original discrete distribution. 
A straightforward remedy would be the finite mixture of regressions, but it still inherits the limitation of the parametric component distribution \citep{canale2017robustifying}. 
Additionally, the specification for the regression function in Step~1 also affects the final outcome of this approach, and it can differ from that in Step~3. 
Therefore, the model specification in this approach is cumbersome. 
Furthermore, the continuous zero-inflated Poisson (cZIP) model considered by \cite{lamarche2021conditional} fails to infer quantiles with levels below the probability of structural zero. 
More precisely, in the implementation of the cZIP approach, one has to solve $\tau=\pi + (1-\pi)F_Y^*(y_\tau^*)$ for $y_\tau^*$, where $\pi$ is the probability of structural zero and $F_Y^*(\cdot)$ is the distribution function of the continuous Poisson defined above. 
However, since a unique $y_\tau^*$ exists only for $\tau>\pi$, the continuous conditional quantiles for lower quantiles are not uniquely determined for many observations. 
Although one may run a regression using only the available $y_\tau^*$'s, this can result in large errors, as demonstrated in the simulation study. 
This would be the reason why \cite{lamarche2021conditional} implemented their approach only for the middle and higher quantiles.

In contrast, the proposed method does not assume a parametric form in $F(y^*|\vx)$, meaning that we do not have to specify a parametric distribution or a functional form for regression in $F(y^*|\vx)$. 
This is because $F(y^*|\vx)$ is derived from a nonparametric mixture for the joint distribution of $y$ and $\vx$ as described in Section~\ref{sec:quantile}, and it is flexible enough to capture the covariate effect on the conditional quantiles well. 
Contrary to \cite{lamarche2021conditional}, the proposed method can be implemented for datasets with excess zeros for any quantile levels. 
This is because the nonparametric mixture with truncated Gaussians creates components with high densities in $y^*\in(-1,0]$ to capture zero inflation without an additional model structure, as in the cZIP model. 
Therefore, the continuous conditional quantiles in the case of zero inflation can be obtained in the same way as in the standard case.

\section{Simulation study}\label{sec:sim}
\subsection{Setting}\label{sec:sim_setting}
Here, we evaluate the performance of the proposed method using simulated datasets. 
We consider the following five settings with $N=1000$ observations. 
In Setting~1, the observations are generated from the Poisson distribution $y_i\sim Poi(\exp(\vx_i'\vbeta))$, where  $\vx_i=(1,x_i)'$, $x_i\sim U(-3,3)$ and $\vbeta=(1,0.5)'$. 
Setting~2 considers the four-component mixture of Poisson and binomial regressions with the mixing weights $(0.2,0.3,0.3,0.2)$. 
The first two components are the Poisson regression models $Poi(\exp(\vx_i'\beta_k))$, $k=1,2,$ and latter two are the binomial $Bin(\text{logit}^{-1}(\vx_i'\beta_k))$,  $k=3,4,$ where $\vx_i=(1,x_i)'$, $x_i\sim N(0,1)$, 
$\vbeta_1=(0.8,0.4)'$,
$\vbeta_2=(2,0.3)'$,
$\vbeta_3=(-1,0.3)'$ and
$\vbeta_4=(-1,-0.1)'$. 
In Setting~3, the pairs of $(y_i^*,x_i)$ are generated from the five-component mixture of the bivariate truncated normal distributions. 
The $k$th component distribution is denoted by $TN_2(\vmu_k,\mSigma_k,\va), k=1,\dots,4$ where $\va=(-1,-\infty)'$,  
$\vmu_1= (1,5)'$, $\vmu_2=(5,7.5)'$, $\vmu_3=(8,10)'$, $\vmu_4=(10,12.5)'$, $\vmu_5=(8,15)'$, 
$\mSigma_1=
\begin{pmatrix}
    2&0.6\\0.6&1
\end{pmatrix}
$, 
$\mSigma_2=
\begin{pmatrix}
    1.5&0.6\\0.6&1
\end{pmatrix}
$,
$
\mSigma_3=
\begin{pmatrix}
    2&0.4\\0.4&1
\end{pmatrix}
$,
$
\mSigma_4=
\begin{pmatrix}
    3&0.4\\0.4&1
\end{pmatrix}
$, and
$
\mSigma_5=
\begin{pmatrix}
    1&-0.3\\-0.3&1
\end{pmatrix}
$. 
The mixing weights are given by $(0.4,0.1,0.2,0.15,0.15)$. 
The latent continuous variable $y^*_i$ is then converted to the count response $y_i$ through \eqref{eqn:group}. 

Settings~4, 5, and 6 are the zero-inflated versions of Settings~1, 2, and 3, respectively. 
Setting~4 considers the zero-inflated Poisson model $y_i\sim \pi_i + (1-\pi_i)Poi(\exp(\vx_i'\vbeta))$ where $\pi_i=\text{logit}^{-1}(\vx_i'\vgamma)$, $\vbeta=(1,0.5)'$ and $\vgamma=(-1.5,-0.5)'$. 
Setting~5 adds a component of $Bin(\text{logit}^{-1}(\vx_i'\vbeta_5))$ to the mixture model of Setting~2 with the uniform component weights.  
This additional component increases the fraction of zeros. 
Finally, Setting~6 also adds a component to the mixture model of Setting~3. 
The parameters of the sixth component are given by $\vmu_6=(-1,5)'$, 
$
\mSigma_6=
\begin{pmatrix}
    0.001&0\\0&1
\end{pmatrix}
$. 
The mixing weights are given by $(0.2,0.15,0.15,0.15,0.15,0.2)$. 
The fractions of zeros of Settings~4, 5, and 6 are $0.31$, $0.18$ and $0.24$, respectively.

Figure~\ref{fig:simdata} of the supplementary material presents the scatter plots of $(y_i,x_i)$ and histograms of $y_i$ of two datasets. 
In the figure, the blue lines represent the true continuous conditional quantiles $y_\tau^{*\text{(true)}}$, and red lines represent the count counterparts $Q^\text{(true)}_y(\tau;\vx)=\lceil y_\tau^{*\text{(true)}}\rceil$ for $\tau=0.1$, $0.5$ and $0.9$. 
The figure illustrates the highly nonlinear relationship between $x$ and $y^*_\tau$, as well as the deviation from the Poisson model, particularly in Settings~3 and 6. 

In this simulation study, the quantiles for counts $Q^\text{(true)}_y(\tau;\vx)$ are the main target quantities. 
This is because there are no unique $y^{*\text{true}}_\tau$'s for the lower quantiles under the ZIP model in Setting~4 due to the same reason described in Section~\ref{sec:parametric}, while the count counterparts are available through the ceiling function.

The prior distributions for the proposed model are specified to ensure that the prior adequately covers the support of the data. 
For the parameters of the base measure of $\mu_y$ and $\vmu_{\vx}$, we set $\mu_{y0}$ and $\vmu_{\vx0}$ equal to the sample means,  $\sigma^2_{y0}$ and $\mSigma_{\vx_0}$ equal to the twice the sample variances. 
For $\sigma^2_y$, $k_0=3$ and we set $t_0$ equal to $k_0-1$ times the sample variance of $y_i$. 
For $\mSigma_c$, $n_0=P+2$, and we set $\mS_0$ equal to $n_0-P-1$ times the diagonal matrix with the sample variance of $\vx_i$ on the diagonal. 
For $\veta$, $\vmu_{\eta0}$ is equal to the $P$ vector, each element of which is the sample covariance between $x_{ij}$ and $y_i$ divided by the standard deviation of $y_i$ and $E_0=10\mI$. 
For the parameters of the PY process, we set $\phi=0.4777$ and $\vartheta=-0.2171$ so that the prior mean and standard deviation of the number of clusters are equal to $20$. 
The MCMC algorithm is run for 20,000 iterations after 2,000 iterations of burn-in period. 
Every 10th draw is retained for the subsequent procedure. 
After fitting the PY mixture model, the continuous conditional quantiles $y^*_{\tau,i}, \ i=1,\dots,N$ are computed as in Section~\ref{sec:quantile}. 
Then $y^*_{\tau,i}$ are regressed on $x_i$ based on the penalized spline model $y_{\tau,i}^*=f(x_i)+\epsilon_i$ by minimizing \eqref{eqn:loss}.

We compare the proposed method with the two existing approaches. 
The first approach uses the parametric model of \cite{lamarche2021conditional} described in Section~\ref{sec:parametric}. 
It is related to our approach because it also computes the continuous conditional quantiles, which are regressed on the covariates to obtain the regression effects. 
The continuous Poisson model is used in the first three settings, and the continuous ZIP model is used in the remaining settings.  
In Setting~1, Step~1 fits the Poisson regression model $y_i\sim Poi(\lambda_i)$ with $\log\lambda_i = \beta_0+\beta_1x_i$, which is precisely the true data-generating model.  
In Settings~2 and 3, Step~1 fits the penalized spline additive Poisson regression model using \texttt{SOP}: $y_i\sim Poi(\lambda_i)$, $\log\lambda_i=f(x_i)$. 
In Settings~4, 5, and 6, Step~1 fits the ZIP model using \texttt{pscl} package. 
In Setting~4, the Poisson mean is modeled as $\lambda_i=\exp(\beta_0+\beta_1x_i)$ and the probability of structural zero is modeled as $\pi_i=\text{logit}^{-1}(\gamma_0+\gamma_1x_i)$. 
In Settings~5 and 6, each regression model includes $x_i^2$ as an additional covariate, since \texttt{pscl} does not support penalized additive regression. 
After calculating the continuous conditional quantiles based on the continuous counterpart, a normal additive regression model is fitted using \texttt{SOP} in all settings. 

The second existing approach is the Bayesian version of the jittering \citep{Machado2005} approach considered by \cite{LeeTereza}. 
This approach smooths count responses differently, by adding a uniform random variable and then using a monotone transformation. 
Thus, this approach considers a quantile regression model for the jittered response given by $\log(y_i+u_i)=\tau+\exp(\vx_i'\vbeta_\tau)+\epsilon_i$, where $u_i\sim U(0,1)$ and $\epsilon_i$ is the error term following the asymmetric Laplace distribution \citep{YuMoyeed}. 
From this, the quantile function for the count response is obtained as $Q_y(\tau;\vx)=\lceil \tau+\exp(\vx'\vbeta_\tau)- 1 \rceil$. 
In Settings~1, 2, 4, and 5, the jittered count response is regressed on the constant term, $x_i$, and its square.  
In Settings~3 and 6, we consider the Bayesian penalized spline version, where the penalty matrix is constructed from the second-order difference matrix \citep[see Section~\ref{sec:generalbayes} and][]{Brezeger}.

\subsection{Results}
Figure~\ref{fig:post_density} of the supplementary material presents the posterior means of the joint densities of $(x_i,y_i^*)$ under the proposed method, obtained from Step~1 of Algorithm~\ref{tab:algo1}. 
The gray circles in the figure indicate the observed pairs $(x_i,y_i)$. 
The density is shown for $y^*>-1$ as the multivariate truncated normal kernel is used. 
The performance of the quantile regression under the proposed method hinges on the estimate of the joint distribution, as the continuous conditional quantiles obtained from this distribution are used in the subsequent steps of the algorithm. 
The figure shows that the model fits the data reasonably well in both settings. 
It should be noted that in the zero-inflated cases of Settings 4, 5, and 6, the nonparametric mixture places high densities in $y^*\in(-1,0]$ to capture the high probabilities of $y=0$.

Table~\ref{tab:sim_comp} presents the averages of the squared errors for $y_\tau^*$ defined by $\text{ASE}=\frac{1}{N}\sum_{i=1}^N(\hat{y}_{\tau,i}^*-y_{\tau,i}^{*(\text{true})})^2$ for the observed data points, where $\hat{y}_{\tau,i}^*$ and $y_{\tau,i}^{*(\text{true})}$, respectively, denote the estimate and true value of $y_{\tau,i}^*$. 
For the proposed method, $\hat{y}_{\tau,i}^*$ is the posterior mean obtained from Step~2 of Algorithm~\ref{tab:algo1}. 
For the continuous Poisson, it is obtained from Step~2 of Algorithm~2 of the supplementary material. 
The table also presents the integrated squared errors (ISE) given by $\text{ISE}=\int \{Q^{(true)}_y(\tau;x)-\hat{Q}_y(\tau;x)\}^2\text{d}x$ for the count quantiles $Q_y(\tau;x)$. 
For the proposed approach, we use the posterior median of $\lceil Q_y^*(\tau;\vx)\rceil$ for $\hat{Q}_y(\tau;x)$ (see Section~\ref{sec:generalbayes}). 
Similarly, for the continuous Poisson and ZIP models, $\hat{Q}_y(\tau;x)=\lceil Q_y^*(\tau;\vx)\rceil$ is obtained using the estimate from Step~3 of Algorithm~\ref{tab:algo2}. 
For the jittering approach, we use the posterior median (see Section~\ref{sec:sim_setting}).

While the proposed method estimated $y_{\tau,i}^*$ well overall, the parametric approach incurred large ASEs for $\tau=0.1$ and $0.9$, except for Settings~1 and 4 where the true models are precisely those of the parametric approach. 
However, due to the limitation described in Section~\ref{sec:parametric}, the continuous ZIP model failed to compute $y_{\tau,i}^*$ for some observations for $\tau=0.1$ and $0.5$ in Settings~4 and 5, and $\tau=0.1$ in Setting~6, where ASEs for the parametric approach are blanked. 
The result indicates that the continuous Poisson model is inappropriate for modeling conditional quantiles as it only models the mean and variance, and is inflexible for capturing tail structures. 
Also, an inference based on the continuous ZIP can fail for lower quantiles. 
The errors for $y_{\tau}^*$ or inability to compute it in this step of Algorithm~2 affect the final step where the quantile regression model is fitted. 

Compared to the parametric approach, the proposed method accurately estimates the continuous conditional quantiles, as they are obtained from the joint distribution estimated in Figure~\ref{fig:post_density}. 
While the zero-inflated situations of Settings~4, 5, and 6 are more challenging, especially Setting~5, in which there are regions where observations generated from this conditional model are sparse, the proposed model, which does not explicitly model the part for structural zeros, can be implemented even for the lower quantiles. 

The table also shows that the proposed approach resulted in the smallest ISE in most cases. 
Specifically, for Setting~1 for all $\tau$ and Setting~4 for $\tau=0.5$ and $0.9$, since the Poisson and zero-inflated Poisson models are the true data generating models, the corresponding parametric continuous approach performed well.

\begin{figure}[H]
    \centering
    \includegraphics[width=0.87\textwidth]{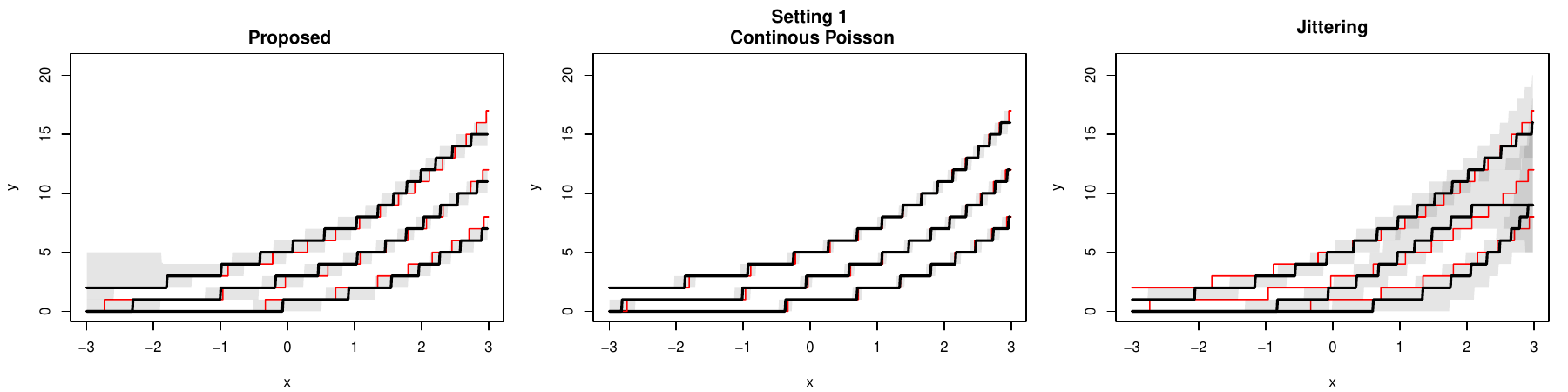}
    \includegraphics[width=0.87\textwidth]{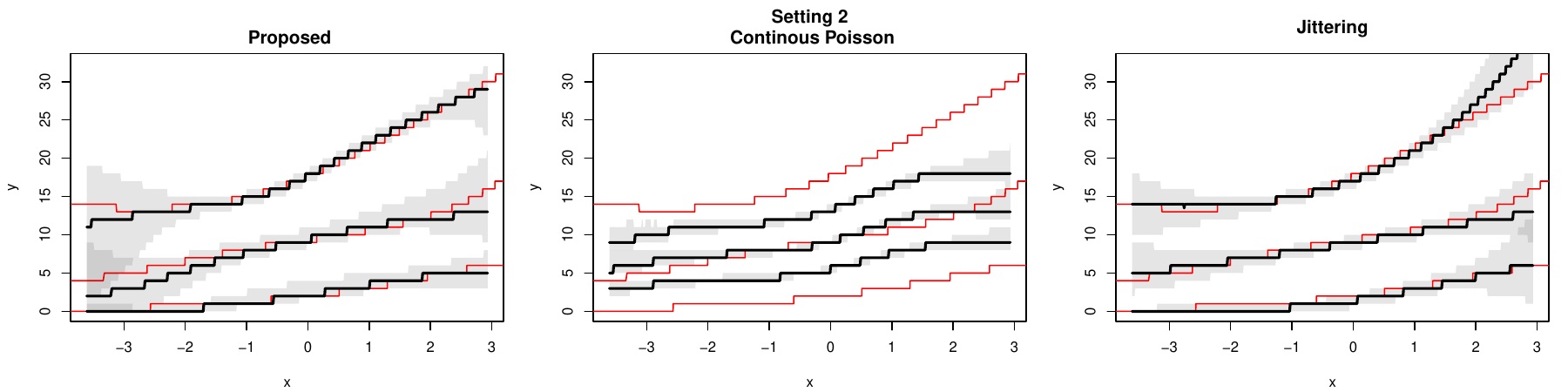}
    \includegraphics[width=0.87\textwidth]{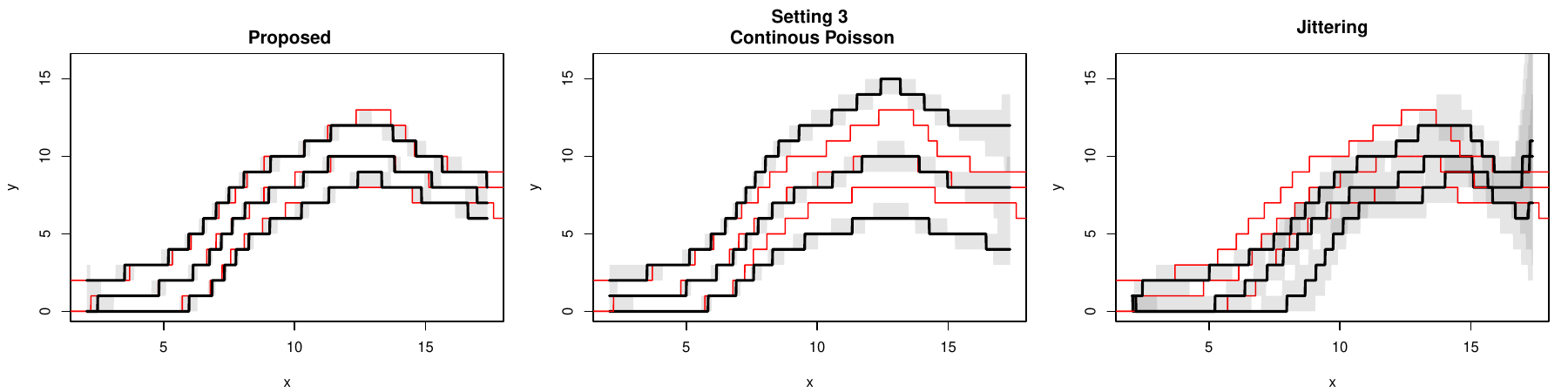}
    \includegraphics[width=0.87\textwidth]{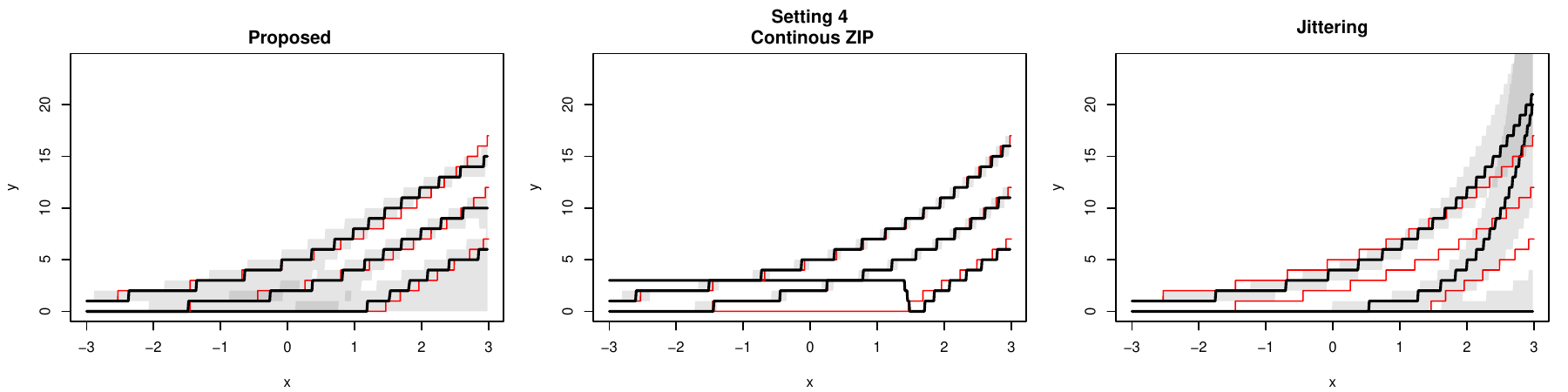}
    \includegraphics[width=0.87\textwidth]{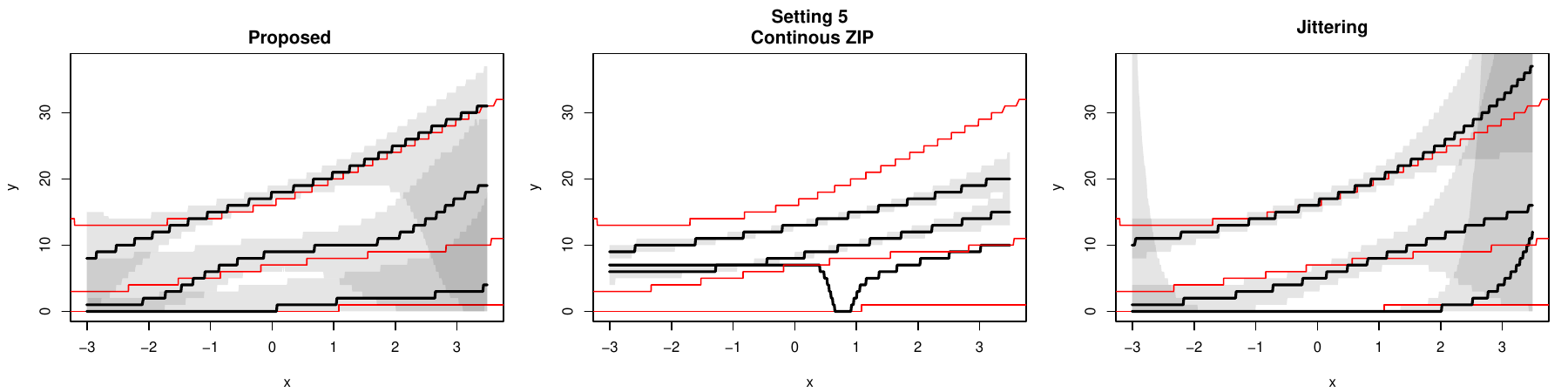}
    \includegraphics[width=0.87\textwidth]{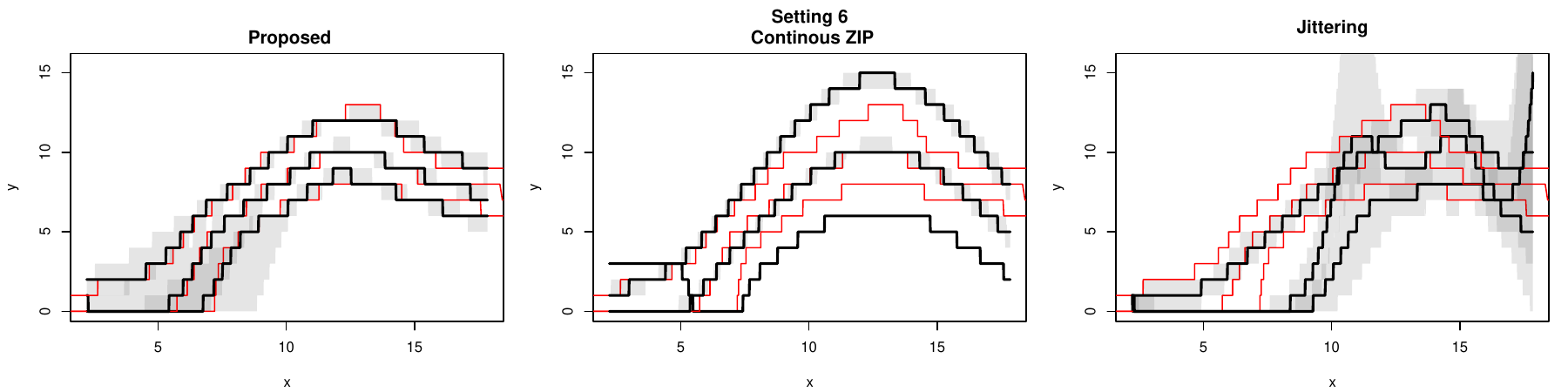}
    \caption{Estimates for count quantiles (solid lines) and 95\% intervals (shaded areas) and true count quantiles (red lines) for $\tau=0.1, 0.5, 0.9$. }
    \label{fig:post_reg}
\end{figure}

Figure~\ref{fig:post_reg} presents the estimates for the count quantiles and associated 95\% intervals, indicated by the gray shades, for the proposed, parametric, and jittering approaches for $\tau=0.1$, $0.5$, and $0.9$. 
The true regression functions are indicated in red. 
For the proposed and jittering methods, the posterior means are used as estimates. 
Since the additive model is used for the proposed model, the regression model closely follows the estimated continuous conditional quantiles $y_{\tau,i}^*$. 
Thus, the count quantiles are well estimated in most cases. 
In Setting~5, the proposed approach resulted in wide credible intervals in the right region due to the sparse observations in this region. 

For the parametric approach, the middle column of the figure shows the bias for the estimated conditional quantile functions except for Setting~1.  
The estimated conditional quantile functions imply that the conditional distribution is estimated to be too dispersed in Setting~2 and too narrow in Setting~3. 
Similar patterns are observed in Settings~5 and 6 for $\tau=0.9$. 
This is the consequence of the errors in estimating $y^*_\tau$ due to the inflexibility of the Poisson assumption. 
Note that in Settings~4, 5, and 6, additive regression models are fitted only to the available $y_{\tau,i}^*$'s. 
The figure also shows the failure of this crude strategy for $\tau=0.1$. 

In the right column of the figure, the jittering approach generally performs poorly, except for Setting~2. 
This approach typically produced wider 95\% intervals than the proposed and continuous Poisson approaches. 
Additionally, in Setting~6, we observe the crossing of the estimated conditional quantiles.

Finally, we also considered an alternative prior specification for the PY process. 
In the alternative prior, we set $\phi=0.6442$ and $\vartheta= -0.5749$ so that the prior mean of the number of clusters remains the same, but the prior standard deviation is doubled. 
The prior specification for the remaining parameters remains the same. 
Figures~\ref{fig:alt_k} and \ref{fig:alt_reg} of the supplementary material, respectively, present the posterior distributions of numbers of clusters and $y_\tau^*$ for $\tau=0.1$, $0.5$, and $0.9$ under the default and alternative prior specifications for the PY process. 
Overall, the results between the two priors are similar. 
In Setting~3, for example, the posterior means of the number of clusters are  $3.780$ and $3.848$ under the default and alternative priors, respectively.
The 95\% credible intervals are $[3,14]$ for both priors. 
In Setting~5, the posterior means are $7.533$ and $6.856$, and the 95\% credible intervals are $[3,14]$ and $[3,13]$ for the default and alternative priors, respectively.

\begin{table}[H]
    \centering
    \caption{Simulation study. Averages of squared errors (ASE) for $y_{\tau,i}^*$ for the observed data points and integrated squared errors (ISE) for $Q_y(\tau;\vx)$ under the proposed, parametric, and jittering approaches.}
    \begin{tabular}{llrrrrr}\toprule
        && \multicolumn{2}{c}{ASE} & \multicolumn{3}{c}{ISE}\\
        \cmidrule(lr){3-4}\cmidrule(lr){5-7}
        &$\tau$& Proposed & Parametric & Proposed & Parametric & Jittering\\\hline
        Setting~1 & 0.1 &0.099 &  0.001 &  1.295 &  0.130 & 2.705\\
                  & 0.5 &0.032 &  0.002 &  1.085 &  0.270 & 6.425\\
                  & 0.9 &0.113 &  0.004 &  1.530 &  0.285 & 3.010\\\hline
        Setting~2 & 0.1 &0.095 & 13.934 &  1.845 & 87.770 & 2.760\\
                  & 0.5 &0.304 &  0.508 & 10.288 &  7.920 & 5.030\\
                  & 0.9 &0.152 & 20.497 &  4.323 &193.955 &17.228\\\hline
        Setting~3 & 0.1 &0.053 &  2.062 &  3.715 & 42.145 &41.860\\
                  & 0.5 &0.031 &  0.051 &  1.955 &  2.365 &31.175\\
                  & 0.9 &0.083 &  1.913 &  3.415 & 42.975 &38.195\\\hline
        Setting~4 & 0.1 &0.578 &  ---   &  1.000 & 40.700 &23.985\\
                  & 0.5 &0.164 &  ---   &  1.320 &  0.310 &28.873\\
                  & 0.9 &0.267 &  0.006 &  2.190 &  0.445 & 9.693\\\hline
        Setting~5 & 0.1 &1.026 &  ---   &  6.395 &292.240 &23.955\\
                  & 0.5 &2.263 &  3.417 & 71.548 & 49.140 &43.585\\
                  & 0.9 &1.181 & 17.119 & 17.510 &205.925 &22.110\\\hline
        Setting~6 & 0.1 &0.236 &  ---   & 4.333  & 84.845 &78.445 \\
                  & 0.5 &0.329 &  ---   & 2.600  & 11.155 &112.455 \\
                  & 0.9 &0.181 & 2.320  & 5.615  & 42.105 & 44.925\\\bottomrule
    \end{tabular}
    \label{tab:sim_comp}
\end{table}

\section{Analysis of length of hospital stay for acute myocardial infarction}\label{sec:empirical}
\subsection{Data and setting}

Using the Worcester Heart Attack Survey (WHAS) data,  available in the R package, \texttt{smoothHR}, we analyze the length of stay (number of days; \texttt{LoS}) for 500 patients hospitalized with acute myocardial infarction. 
The data also includes information on patients, which we use as covariates. 
They are the age of the patient (\texttt{age}), heart rate on the first day of hospitalization (\texttt{hr}), body mass index (\texttt{bmi}), diastolic blood pressure (\texttt{distbp}) and systolic blood pressure (\texttt{sysbp}). 
Figure~\ref{fig:whas} of the supplementary material displays the scatter plots of \texttt{LoS} and covariates and histogram of \texttt{LoS}.
In the scatter plots, while the lower conditional quantiles of \texttt{LoS} may not change with the covariate values, the covariates may have nonlinear effects on the upper conditional quantiles. 
Therefore, the relationship between the conditional quantiles of \texttt{LoS} and covariates may vary across quantiles. 
The proposed method is expected to reveal the varying effects of the covariates on the conditional quantiles of length of stay, providing a deeper insight into patient outcomes. 
Table~\ref{tab:summary} of the supplementary material presents the sample averages and standard deviations of the variables.

We fit the proposed model to estimate the six-dimensional joint distribution and obtain the continuous conditional distribution and its quantiles $y^*_\tau$ of \texttt{LoS} given the covariates. 
The same prior setting is used except for $\phi=0.6036$ and $\vartheta=-0.5322$ for the PY process, such that the prior mean and standard deviation of the number of clusters are 10 and 20, respectively. 
As in the simulation study, using \texttt{SOP}, the additive model regresses the continuous conditional quantiles on the covariates: 
$
y^*_{\tau,i}=\text{constant}_\tau+f_{\tau,1}(\texttt{age}_i)+f_{\tau,2}(\hr_i)+f_{\tau,3}(\bmi_i)+f_{\tau,4}(\sysbp_i)+f_{\tau,5}(\diasbp_i)+\epsilon_i$, 
where each covariate effect $f_{\tau,j}(\cdot)$ is presented using the B-spline as in Section~\ref{sec:generalbayes}.

The proposed method is compared with the method of \cite{lamarche2021conditional} using the Poisson additive model, where each covariate has a nonlinear effect: 
$y_i|\vx_i\sim Poi(\lambda_i)$, $\log\lambda_i=\text{const.}+g_1(\texttt{age}_i)+g_2(\hr_i)+g_3(\bmi_i)+g_4(\sysbp_i)+g_5(\diasbp_i)$, 
where each covariate effect $g_{j}(\cdot)$ is presented using the B-spline. 
As in the simulation study,  \texttt{SOP} is used to estimate the penalized spline model. 
Then, the same additive model as the proposed method is fitted to obtain the covariate effects on the continuous conditional quantiles. 

Two observations with outlying values for \texttt{sysbp} and \texttt{hr} were removed from the original dataset for the computational stability of the continuous Poisson approach, and the remaining $N=498$ observations were used to implement both models. 
For both models, all covariates are standardized before the model fitting. 

For model comparison, we investigate the performance of the two approaches in predicting conditional quantiles of \texttt{LoS}. 
We randomly split the data into the training and test sets. 
The models are fitted to the training set, which consists of four-fifths of the observations, and then the conditional quantiles of \texttt{LoS} in the test set are predicted. 
The predictive performance for the conditional quantiles is measured using the quantile-weighted continuous rank probability score (CRPS) of \cite{Gneiting01072011}, defined by
\begin{equation}
    \text{CRPS}(Q_\tau,y^\text{test})=\int_0^1 2\left(I\left\{y^\text{test}\leq \hat{Q}_y(\tau;\vx)\right\}-\tau\right)(\hat{Q}_y(\tau;\vx) - y^\text{test})\nu(\tau)\d\tau,
\end{equation}
where $y^\text{test}$ corresponds to the value of \texttt{LoS} in the test set, $I(\cdot)$ denotes the indicator function, $\nu(\tau)$ is the weight function, and $\hat{Q}(\tau;\vx)$ is the predicted $\tau$th conditional quantile of \texttt{LoS} given the covariates in the test set. 
For the proposed approach, $\hat{Q}(\tau;\vx)$ is the posterior predictive median of $Q_y(\tau;\vx)$ for the test set. 
For the continuous Poisson, $\hat{Q}(\tau;\vx)$ is computed from the regression model fitted in Step~3 of Algorithm~\ref{tab:algo2}. 
The integral is approximated using the trapezoid rule by predicting the conditional quantiles for $\tau=0.1,\dots,0.9$. 
The smaller CRPS indicates a better conditional quantile prediction. 
Random splits are repeated independently twenty times. 
For each split, the CRPS is averaged over the test-set observations for each model. 
For the weight function $\nu(\tau)$, we consider three types, $\nu(\tau)=1$, $\tau^2$, and $(1-\tau)^2$, respectively, labeled as `none', `right', and `left'. 
The latter two emphasize the right tail and left tail, respectively.

\subsection{Results}
\begin{figure}
    \centering
    \includegraphics[width=0.5\linewidth]{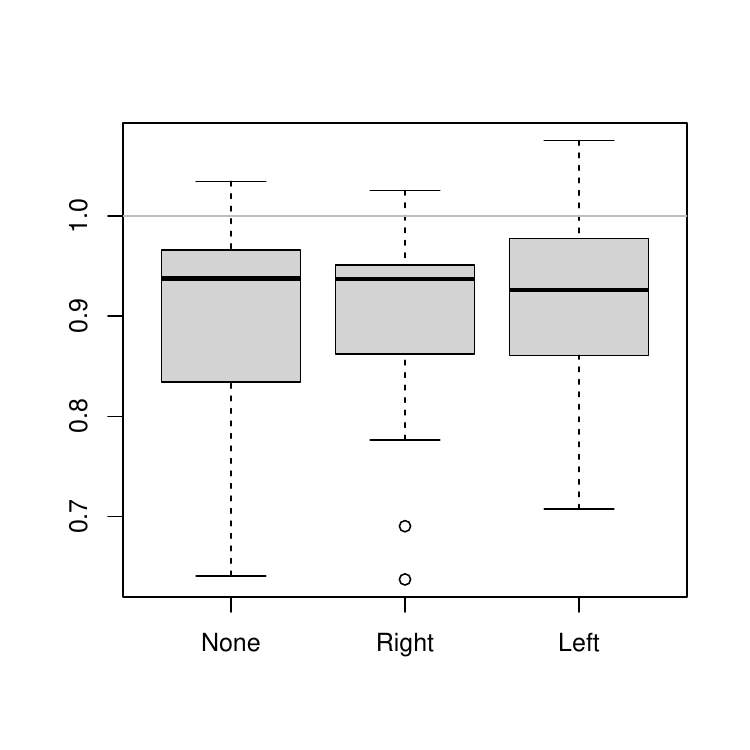}
    \caption{Ratio of CRPS for the proposed approach to that for the continuous Poisson over the twenty independent splits of the data. Values lower than $1$ indicate the superiority of the proposed approach. }
    \label{fig:crps}
\end{figure}

First, Figure~\ref{fig:crps} presents the boxplots of the ratio of CRPS for the proposed method to that for the continuous Poisson over the twenty splits of the data. 
The figure shows that the CRPS ratios are mostly below one, indicating the superior performance of the proposed method in predicting the conditional quantiles of \texttt{LoS}. 

Next, we compare the continuous conditional quantiles of \texttt{LoS} estimated by the proposed and continuous Poisson approaches using the entire data. 
The left column of Figure~\ref{fig:sp-ytau} of the supplementary material presents the posterior means of $y_\tau^*$ obtained from Step~1 in Algorithm~\ref{tab:algo1} given the observed covariates for $\tau=0.1$, $0.5$, and $0.9$. 
The relationship between $y_\tau^*$ and each covariate varies for the different quantiles under the proposed model. 
Notably, the estimated continuous conditional quantiles for $\tau=0.9$ exhibit considerable variation. 
In contrast, in the right column of Figure~\ref{fig:sp-ytau}, which shows the estimates of $y_\tau^*$ and covariates obtained from Step~2 of Algorithm~2, the relationships between the estimated $y_\tau^*$ and the covariates do not vary much between the different quantile levels except for the levels of $y_\tau^*$. 
It is important to note that the parametric limitation and inflexibility in this stage can impact the regression result, as highlighted in the simulation study. 

Figure~\ref{fig:trace} presents the trace plots of $f_{\tau,j}(x_j)=\mB(x_{j})'\vbeta_{\tau,j}$ evaluated at arbitrary points for $\tau=0.1$, $0.5$, and $0.9$, computed from the output of Algorithm~\ref{tab:algo1} of the proposed method. \label{page:trace}
The figure shows the convergence of the algorithm. 

\begin{figure}
    \centering
    \includegraphics[height=\textheight]{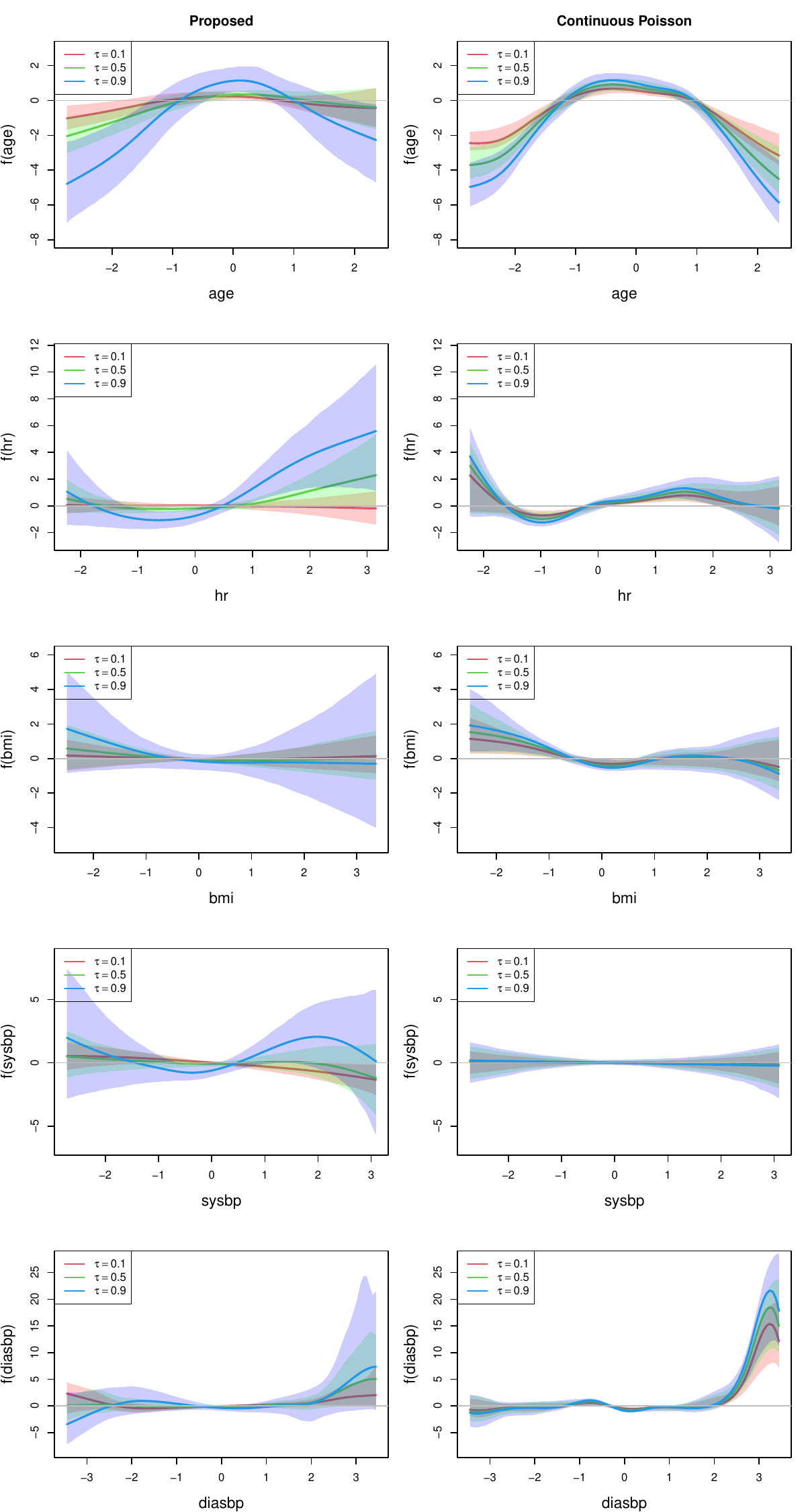}
    \caption{Posterior means (solid lines) and 95\% intervals (shaded areas) of the covariate effects $f(x_j)$ for the proposed (left) and continuous Poisson (right) approaches for $\tau=0.1$, $0.5$, and $0.9$. }
    \label{fig:reg-prop}
\end{figure}

Figure~\ref{fig:reg-prop}  presents the posterior means and 95\% credible intervals of the function $f_{\tau,j}(x_j)$ for $\tau=0.1$, $0.5$ and $0.9$ under the proposed and continuous Poisson approaches. 
In each panel, the estimated functions of each covariate for the three quantiles are overlaid, and the shaded areas represent the 95\% intervals. 
As expected from Figure~\ref{fig:sp-ytau}, the effects of covariates on the continuous conditional quantiles vary flexibly across the different quantiles. 
This is particularly true for \texttt{age} and \texttt{hr}. 
For $\tau=0.1$ and $0.5$, younger patients with ages below the average have shorter lengths of stay, while the age of patients above the average does not influence these quantiles of \texttt{los} as the 95\%  intervals include zero. 
For example, the posterior mean and 95\%  interval for $f_1(-2.023)$, which corresponds to 40.46 years of age, are, respectively, $-0.593$ and $(-1.109, -0.056)$ for $\tau=0.1$ and $-1.283$ and $(-2.023, -0.586)$ for $\tau=0.5$. 
For $\tau=0.9$, the effect of \texttt{age} is more pronounced, significantly influencing the upper quantiles of \texttt{LoS}. 
For the upper quantiles of \texttt{LoS}, patients of both younger and older ages have a longer stay than those of the average age. 
In this case, the posterior mean and 95\%  interval of $f_1(-2.023)$ are, respectively, $-3.281$ and $(-5.206, -1.410)$. 
On the other hand, those of $f_1(-0.014)$, which corresponds to 69.62 years of age and is close to the average age, are $1.121$ and $(0.471,1.892)$, respectively. 
For $\tau=0.1$ and $0.5$, the posterior means of $f_1(-0.014)$ are $0.239$ and $0.327$, respectively, and 95\% intervals are $(0.059, 0.450)$ and $(0.136, 0.539)$, respectively. 

The heart rate has little effect on the length of stay for $\tau=0.1$ as the 95\%  interval includes zero in most of the region. 
For $\tau=0.5$, while the interval still includes zero in most regions, the faster heart rate may increase the median length of stay. 
In the case of $\tau=0.9$, on the other hand, the heart rate around the average or below decreases the upper quantile of \texttt{LoS}. 
For example, the posterior mean and 95\%  interval of $f_2(-0.004)$, which corresponds to $86.77$ beats per minute and is close to the average, are $-0.748$ and $(-1.277, -0.202)$. 
On the other hand, higher heart rates are associated with longer hospital stays. 
The posterior mean and 95\% interval of $f_2(2.010)$, which corresponds to $133.48$ beats per minute, are $3.786$ and $(1.335, 6.272)$. 
The remaining panels of Figure~\ref{fig:reg-prop} show that the effects of \texttt{bmi}, \texttt{sysbp}, and \texttt{diasbp} are not confirmed for all three quantiles as the 95\% intervals include zero in most regions. 

Finally, the right column of Figure~\ref{fig:reg-prop} presents the regression result under the continuous Poisson approach. 
As expected from Figure~\ref{fig:sp-ytau}, the shapes of the estimated functions are similar across the quantiles for all covariates, and the 95\% intervals appear to be narrower than those for the proposed method.

\section{Concluding remarks}\label{sec:conc}
We have introduced a novel general Bayesian approach to quantile regression for count data through generative modeling based on a Bayesian nonparametric mixture for the joint distribution of the count response and covariate. 
The proposed method improves the accuracy of quantile estimates and enhances the interpretability of the relationship between the response and explanatory variables, making it a valuable tool for researchers and practitioners working with count data.
It is also notable that the proposed method also works for zero-inflated count data without an additional structure for excess zeros. 
In summary, this paper presents a significant advancement in quantile regression for count data, offering a robust and flexible approach that addresses the limitations of existing methods.

The simulation study and WHAS data analysis demonstrate the efficacy of the proposed method. 
Notably, the WHAS data analysis using the proposed method has revealed that the relationships between the length of stay and covariates vary across quantiles. 
In contrast, the parametric count model, in which the first step regression only models certain parts of the conditional distribution of count responses, failed to capture such varying covariate effects on quantiles.   

The present approach can be extended in various directions as follows. 
This paper considered the quantile regression with generative modeling only for cross-sectional data, while \cite{lamarche2021conditional} considered the parametric approach in the longitudinal setting. 
The proposed model can be extended for longitudinal count data by adopting hierarchical models for Steps~1 and 3 of Algorithm~\ref{tab:algo1}. 
There are various ways to incorporate dependencies into a hierarchical nonparametric mixture model \citep[e.g.,][]{Quintana02072016,NDP, HDP}, and an appropriate choice of a nonparametric prior is typically application-dependent. 
Additionally, extending the regression model in Step~3 to a hierarchical model also expands the design options. 
Quantile regression for ordered categorical data has been recently studied  \citep[e.g.,][]{rahman}. 
The present approach for count responses can also be extended to handle ordered categorical responses.  
In this case, as in \cite{DeYoreo02012018}, the thresholds in the grouping mechanism \eqref{eqn:group} become unknown parameters to be estimated. 
Another important direction of research would be to address sparsity and high-dimensional covariates, which would require introducing sparsity via shrinkage priors or variable selection in Step~1, rather than simply introducing penalized estimation in Step~3. 
Since the primary focus of this paper is to develop a flexible modeling approach for quantile regression of counts, a discussion of high-dimensionality is deemed beyond the scope of this paper. 
Again, there would be various possibilities for such an extension, and the choice depends on the purpose of the analysis. 
We would like to explore these topics in future research.

\section*{Acknowledgements}
This work was supported by JSPS KAKENHI Grant Numbers JP20H00080, JP21H00699, JP21K01421, JP22K13376, JP24K00244, 25H00546.

\begin{appendices}
\numberwithin{figure}{section}
\numberwithin{table}{section}
\numberwithin{equation}{section}
\section{Gibbs sampler with importance conditional sampling}\label{sec:mcmc}
Step~1 of Algorithm~\ref{tab:algo1} requires fitting the nonparametric mixture model using the Gibbs sampler described here. 
Our Gibbs sampler relies on the importance conditional sampling (ICS) algorithm proposed by \cite{canale22}. 
ICS can be easily implemented and is known to perform more stably than the slice sampler \citep{WALKER, KALLI}, which is often used to estimate Bayesian nonparametric mixture models. 

Suppose $\vtheta_1,\dots,\vtheta_n|G\sim G$ and $G\sim PY(\phi,\eta;G_0)$. 
Let $(\vtheta_1^*,\dots,\theta_{K_N}^*)$ and $(n_1,\dots,n_k)$ denote the set of $K_N$ distinct values and corresponding frequencies in $(\vtheta_1,\dots,\vtheta_n)$. 
ICS is based on the result of \cite{pitman1997two} that the conditional distribution of $G$ given $(\vtheta_1,\dots,\vtheta_n)$ can be expressed as the mixture of a PY process $\tilde{G}$ and $K_N$ fixed discrete points. 
\[
\pi_0\tilde{G} + \sum_{k=1}^{K_N}\pi_k\delta_{\vtheta_k^*}(\cdot)
\]
where $(\pi_0,\pi_1,\dots,\pi_{K_N})\sim Dir(\eta+K_N\phi,n_1-\phi,\dots,n_k-\phi)$ and $\tilde{G}\sim PY(\phi,\eta+K_N\phi;G_0)$ is independent of $(\pi_0,\pi_1,\dots,\pi_{K_N})$ \citep[see also][]{canale22}. 
From this, using the notation of the kernel in \eqref{eqn:kernelmix} of the main text, the full conditional distribution of $\vtheta_i$ given data is proportional to
\begin{equation}\label{eqn:pyfcd}
\pi_0k(\vz_i;\vtheta)\tilde{G}(d\vtheta)+\sum_{k=1}^{K_N}\pi_kk(\vz_i;\vtheta_j^*)\delta_{\vtheta_k^*}(d\vtheta), 
\end{equation}
where $\tilde{G}$ is the restriction of $G$ to  $\mTheta\backslash \left\{\vtheta_1^*,\dots,\vtheta_{K_N}\right\}$. 
Note that it is possible to sample $\vvartheta_h|\tilde{G}$, $h=1,\dots,M$ for any $M\geq 1$. 
Throughout this paper, we use $M=10$. 
Furthermore, the generated values have ties with $R_M$ distinct values $(\vvartheta_1^*,\dots,\vvartheta_{R_M}^*)$ associated with the frequencies $(m_1,\dots,m_{R_M})$, because $\tilde{G}$ is  almost surely discrete.  
Then the full conditional distribution \eqref{eqn:pyfcd} can be rewritten as 
\[
\pi_0\sum_{h=1}^{R_M}\frac{m_h}{M}k(\vz_i;\vvartheta_h^*)\delta_{\vvartheta_h^*}(d\vtheta)+\sum_{k=1}^{K_N}\pi_kk(\vz_i;\vtheta_k^*)\delta_{\vtheta_k^*}(d\vtheta), 
\]

Using the parenthesized superscript to denote the value in an algorithm iteration, $s$th iteration of the ICS algorithm proceeds as follows. 
\begin{enumerate}
    \item 
    Sample $(\pi_0^{s},\pi_1^{s},\dots,\pi_{K_N}^{s})$ from $Dir(\eta+\phi K_N^{(s-1)},n_1^{(s-1)}-\phi,\dots,n_{K_N}^{(s-1)}-\phi)$. 
    \item 
    Sample $(\vvartheta_1^{(s)}\dots,\vvartheta_{M}^{(s)})$ based on the following urn scheme:
    \[
    P(\vvartheta_{h+1}^{(s)}\in\cdot|\vvartheta_1^{(s)},\dots,\vvartheta_h^{(s)})=\frac{\eta+\phi(K_N^{(s-1)}+r_h^{(s)})}{\eta+\phi K_N^{(s-1)}+h} G_0(\cdot) 
    + \sum_{j=1}^{r_h^{(s)}}\frac{m_j^{(s)}-\phi}{\eta+\phi K_N^{(s-1)}+h}\delta_{\vvartheta_j^{*(s)}}(\cdot),
    \]
    for $h=0,1,\dots,M-1$, where $r_h^{(s)}$ is the number of distinct values in $(\vvartheta_1^{(s)},\dots,\vvartheta_h^{(s)})$. 
    \item 
    For $i=1,\dots,n$, assign cluster  by sampling $\theta_i^{(s)}$ from the following categorical distribution 
    \[
    P(\vtheta_i^{(s)}=\vtheta|-)\propto\left\{
    \begin{array}{ll}
         \pi_0^{(s)}\frac{m_h^{(s)}}{m}k(\vz_i;\vvartheta_h^{*(s)})& \text{if}\quad \vtheta\in\{\vvartheta_1^{*(s)},\dots,\vvartheta_{R_M^{(s)}}^{*(s)}\} \\
         \pi_k^{(s)}k(\vz_i;\vtheta_k^{*(s)})&\text{if}\quad \vtheta\in \{\vtheta_1^{*(s)},\dots, \vtheta_{K_N^{(s)}}^{*(s)}\}\\
         0&\text{otherwise}. 
    \end{array}
    \right.
    \]
\end{enumerate}

Given the cluster assignment, the cluster parameters $\vtheta_1^{*(s)},\dots,\vtheta_{{K_N}^{(s)}}^{*(s)}$ and latent continuous responses $y_i^{*(s)}$ are updated, where $\vtheta_k^{*(s)}=(\mu_{y,k}^{*(s)},\sigma_{y,k}^{*2(s)},\vmu_{\vx,k}^{*(s)},\mSigma_{c,k}^{*(s)},\veta_k^{*(s)})$ for $k=1,\dots,K_N^{(s)}$. 
Let us denote the set of indices by $C_k^{(s)}=\{i:\vtheta_i^{(s)}=\vtheta_k^{*(s)}\}$. 
For notational simplicity, we drop the parenthesized superscript in what follows. 
We utilize the expression for the kernel mixture \eqref{eqn:kernelmix2} of the main text and alternately update $\mu_{y,k}^*$,  $\sigma_{y,k}^{*2}$, $\vmu_{x,k}^*$, $\mSigma_{c,k}^*$ and $\veta^*_k$. 

\paragraph{Sampling $\mu_{y,k}^*$ and $\sigma_{y,k}^{*2}$:}
We sample $\mu_{y,k}^*$ and $\sigma_{y,k}^{*2}$ in one block.  
From \eqref{eqn:kernelmix2}, the full conditional distribution of $(\mu_{y,k}^*, \sigma_{y,k}^{*2})$ is given by
\[
\begin{split}
&p(\mu_{y,k}^*, \sigma_{y,k}^{*2}|-)\\
&\propto \left(\prod_{i\in C_k}k_{y^*}(y_i^*;\mu_{y,k}^*,\sigma_{y,k}^{*2}) \right)p(\mu_{y,k}^*|\mu_{y0},\sigma_{y0}^2)p(\sigma_{y,k}^{*2}|k_0,t_0)\\
&\propto \left(\prod_{i:\in C_k}\frac{\phi(y_i^*;\mu_{y,k}^*,\sigma_{y,k}^{*2})}{1-\Phi\left(\frac{-1-\mu_{y,k}^*}{\sigma_{y,k}^*}\right)} \right)
\exp\left\{-\frac{(\mu_{y,k}^*-\mu_{y0})^2)}{2\sigma_{y0}^2}\right\}
\left(\frac{1}{\sigma_{y,k}^{*2}}\right)^{k_0+1}\exp\left\{-\frac{t_0}{\sigma_{y,k}^{*2}}\right\}. 
\end{split}
\]
The target distribution is two-dimensional, which is relatively easy to simulate using a standard sampling algorithm. 
We consider a combination of the random walk Metropolis-Hastings (RWMH) algorithm and acceptance-rejection MH (ARMH) algorithm. 
While RWMH is easy to implement, it requires adjusting the step size of the random walk, which is highly cumbersome to a Bayesian nonparametric mixture model. 
ARMH uses the normal approximation around the mode of the target distribution to construct a proposal distribution. 
However, in our experience with the proposed model, the normal approximation can fail in the early stage of the Gibbs sampler. 
Therefore, we begin the sampling with RWMH using a fixed step size and then switch to ARMH after some time, when the observations have been roughly clustered.

\paragraph{Sampling $\vmu_{\vx, k}^*$:}
Given the normal base measure $N(\vmu_{\vx0},\mSigma_{\vx0})$ for $\vmu_{\vx}$, from \eqref{eqn:kernelmix2} of the main text, the full conditional distribution of $\vmu_{\vx, k}^*$ is given by
\[
\begin{split}
    p(\vmu_{\vx, k}^*|-)&\propto \left(\prod_{i\in C_k}k_{\vx;y^*}(\vx_i|y^*_i;\vmu_{\vx, k},\mSigma_{c,k},\veta)\right) p(\vmu_{\vx, k}^*|\vmu_{\vx 0},\mSigma_{\vx 0})\\
    &\propto\left(\prod_{i\in C_k}\exp\left\{-\frac{1}{2}(\vx_i-\vmu_{\vx, k}^*-\sigma_{y,k}^{*-2}\veta_k^*(y_i^*-\mu_{y,k}^*))'\mSigma_{c,k}^{*-1}(\vx_i-\vmu_{\vx, k}^*-\sigma_{y,k}^{*-2}\veta_k^*(y_i^*-\mu_{y,k}^*))\right\}\right)\\
    &\quad\times \exp\left\{-\frac{1}{2}(\vmu_{\vx, k}^*-\vmu_{\vx 0})'\mSigma_{\vx 0}^{-1}(\vmu_{\vx, k}^*-\vmu_{\vx 0})\right\}.
\end{split}
\]
Therefore, for $k=1,\dots,K_N$, $\vmu_{\vx, k}^*$ is sampled from $N(\vm_{\vx, k},\mM_{\vx, k})$ where
\[
\mM_{\vx, k} = \left[n_k\mSigma_{c,k}^{*-1} + \mSigma_{\vx0}^{-1}\right]^{-1},\quad 
\vm_{\vx, k}=\mM_{\vx, k}\left[\mSigma_{c,k}^{*-1}\sum_{i\in C_k}(\vx_i-\sigma_{y,k}^{*-2}\veta_k^*(y_i^*-\mu_{y,k}^*)) + \mSigma_{\vx 0}^{-1}\vmu_{\vx 0}\right]. 
\]

\paragraph{Sampling $\mSigma_{ck}^*$:}
Similarly, with the inverse Wishart base measure $IW(n_0,\mS_0)$, the full conditional distribution of $\mSigma_{ck}^*$ is $IW(n_0+n_k, \mS_{k})$ where
\[
\mS_k = \mS_0 + \sum_{i\in C_k}\ve_i\ve_i',\quad \ve_i=\vx_i-\vmu_{\vx k}^*-\sigma_{y,k}^{*-2}\veta_k^*(y_i^*-\mu_{y,k}^*). 
\]

\paragraph{Sampling $\veta_k^*$:}
With the normal base measure $N(\vmu_{\eta 0},\mSigma_{\eta 0})$, the full conditional distribution of $\veta_k^*$ is $N(\vb_k,\mB_k)$ where
\[
\begin{split}
\mB_k &= \left[\sigma_{y,k}^{*-4}\sum_{i\in C_k}(y_i^*-\mu_{y,k}^*)^2 \mSigma_{c,k}^{*-1}+\mSigma_{\eta0}^{-1} \right]^{-1},\\
\vb_k &= \mB_k\left[\sigma_{y,k}^{*-2}\sum_{i\in C_k} (y_i^*-\mu_{y,k}^*)\mSigma_{c,k}^{*-1}(\vx_i-\vmu_{\vx, k}^*)+ \mSigma_{\eta0}^{-1}\vmu_{\veta0}\right]. 
\end{split}
\]

\paragraph{Sampling $y_i^*$:}
For $i=1,\dots,N$, the latent continuous response $y_i^*$ associated with $y_i=g$ is sampled from $N(m_i,v^2_i)$ truncated on the interval $(g-1,g]$, where $m_i=\mu_{y,i}+\veta_i\mSigma_{c,i}^{-1}(\vx_i-\vmu_{\vx,i})$ and $v^2_i=\sigma_{y,i}^2-\veta'\mSigma_{c,i}^{-1}\veta$.

\section{Computing continuous quantiles}\label{sec:cq}
The details of Step~2 of Algorithm~\ref{tab:algo1} are described here. 
Given $s$th posterior draw of the parameters, from \eqref{eqn:pyfcd}, the draw of the conditional density is given by
\[
\begin{split}
f(y^*|\vx;G^{(s)}) &=
\frac{\int k(\vz;\vtheta)\d G^{(s)}(\vtheta)}{\int k_\vx(\vx;\vtheta_\vx)\d G^{(s)}(\vtheta_\vx)}\\
&=\frac{1}{D^{(s)}}\left(\pi_0^{(s)}\sum_{h=1}^{R_M^{(s)}}\frac{m_h^{(s)}}{M} k(\vz;\vvartheta_h^{*(s)}) + \sum_{k=1}^{K_N^{(s)}}\pi_k^{(s)}k(\vz;\vtheta_k^{(s)})\right)\\
D^{(s)}&=\pi_0^{(s)}\sum_{h=1}^{R_M^{(s)}}\frac{m_h^{(s)}}{M} k_\vx(\vx;\vvartheta_{\vx,h}^{*(s)}) + \sum_{k=1}^{K_N^{(s)}}\pi_k^{(s)}k_\vx(\vx;\vtheta_{\vx,k}^{(s)}). 
\end{split}
\]
Since $k(\vz;\vtheta)=k_{y^*|\vx}(y^*|\vx;\vtheta_{y^*|\vx}) k_\vx(\vx;\vtheta_\vx)$ where $k_{y^*|\vx}$ is the conditional kernel of $y^*$ given $\vx$, which is $N(\mu_{y^*|\vx},\sigma_{y^*|\vx}^2)$ truncated on $y^*>-1$, and $\vtheta_{y^*|\vx}=(\vmu_{y^*|\vx},\sigma_{y^*|\vx}^2)$, $s$th draw of the conditional distribution function is given by
\[
\begin{split}
F(y^*|\vx;G^{(s)})&=\int_{-1}^{y^*}\frac{1}{D^{(s)}}
\biggl(\pi_0^{(s)}\sum_{h=1}^{R_M^{(s)}}\frac{m_h^{(s)}}{M} k_{y^*|\vx}(t^*|\vx;\vvartheta_{y^*|\vx,h}^{(s)})k_\vx(\vx;\vvartheta_{\vx,h}^{(s)})\\
&\quad\quad+\sum_{k=1}^{K_N^{(s)}}\pi_k^{(s)}k_{y^*|\vx}(t^*|\vx;\vvartheta_{y^*|\vx,h}^{(s)})k_\vx(\vx;\vvartheta_{\vx,h}^{(s)})\biggr)\d t^*\\
&=\frac{1}{D^{(s)}}\biggl(\pi_0^{(s)}\sum_{h=1}^{R_M^{(s)}}\frac{m_h^{(s)}}{M} K_{y^*|\vx}(y^*|\vx;\vvartheta_{y^*|\vx,h}^{(s)})k_\vx(\vx;\vvartheta_{\vx,h}^{(s)})\\
&\quad\quad+\sum_{k=1}^{K_N^{(s)}}\pi_k^{(s)}K_{y^*|\vx}(y^*|\vx;\vvartheta_{y^*|\vx,k}^{(s)})k_\vx(\vx;\vvartheta_{\vx,k}^{(s)})\biggr)
\end{split}
\]
where $\vvartheta_{y^*|\vx}$ denotes the draw of $\vtheta_{y^*|\vx}$ from $\tilde{G}$ and denoting the standard normal distribution function by $\Phi(\cdot)$ and
\[
K_{y^*|\vx}(y^*|\vx;\vtheta_{y^*|\vx})= \frac{\Phi\left(\frac{y^*-\mu_{y^*|\vx}}{\sigma_{y^*|\vx}}\right)-\Phi\left(\frac{-1-\mu_{y^*|\vx}}{\sigma_{y^*|\vx}}\right)}{1-\Phi\left(\frac{-1-\mu_{y^*|\vx}}{\sigma_{y^*|\vx}}\right)}. 
\]
Then using the above expressions,  $s$th posterior draw of the continuous quantile $y_{\tau,i}^{*(s)}$ is obtained by finding a value which satisfies 
$F(y_{\tau,i}^{*(s)} | \vx_i;G^{(s)})= \tau$. 
This equation is solved numerically.

\section{Algorithm for the parametric approach using the continuous counterpart}\label{sec:algo2}

\setcounter{algorithm}{1}
\begin{algorithm}[H]
    \caption{Parametric approach using the continuous counterpart}
    \begin{algorithmic}[1]
    \Require {Count response $y_i$ and covariate $\vx_i$ for $i=1,\dots,N$. }
    \item
    Estimate the usual discrete regression model and obtain the model estimate $\hat{\vtheta}$. 
    \item 
    Compute the conditional quantile of the continuous version of the model $y_{\tau,i}^*$ for $i=1,\dots,N$, by fixing the parameter value to $\hat{\vtheta}$. 
    \item 
    Regress $y_{\tau,i}^*$ on $\vx_i$ to find the quantile-specific regression parameter $\hat{\vbeta}_\tau$ by minimizing the loss function such as \eqref{eqn:loss} of the main text. 
    \item 
    The standard errors and confidence intervals are constructed using bootstrap. 
    \Ensure The estimate of the regression parameter $\hat{\vbeta}_\tau$ and the bootstrap sample $\vbeta_\tau^{(s)}$ for $s=1,\dots,S$. 
    \end{algorithmic}
    \label{tab:algo2}
\end{algorithm}

\section{Additional figures and tables for the simulation study and real data analysis}
\label{sec:fig}

\begin{figure}[H]
\centering
		\includegraphics[scale=0.5]{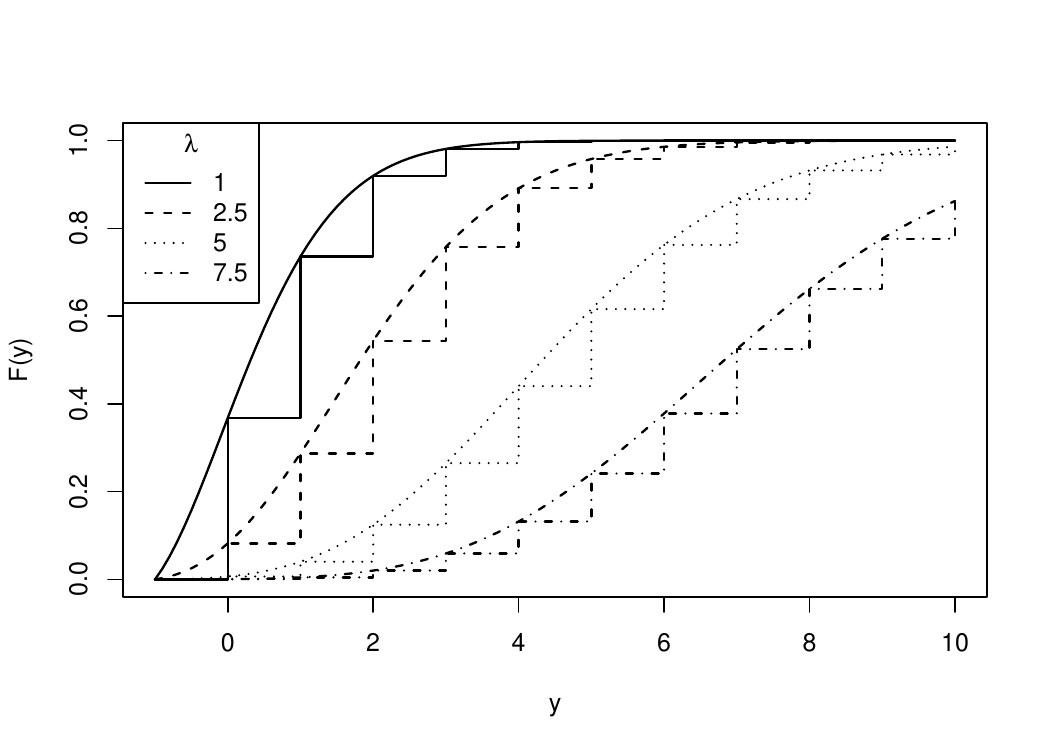}
		\caption{The distribution functions of the continuous and discrete Poisson distributions.}
        \label{fig:poi}
\end{figure}

\begin{figure}[H]
    \centering
    \includegraphics[width=0.32\textwidth]{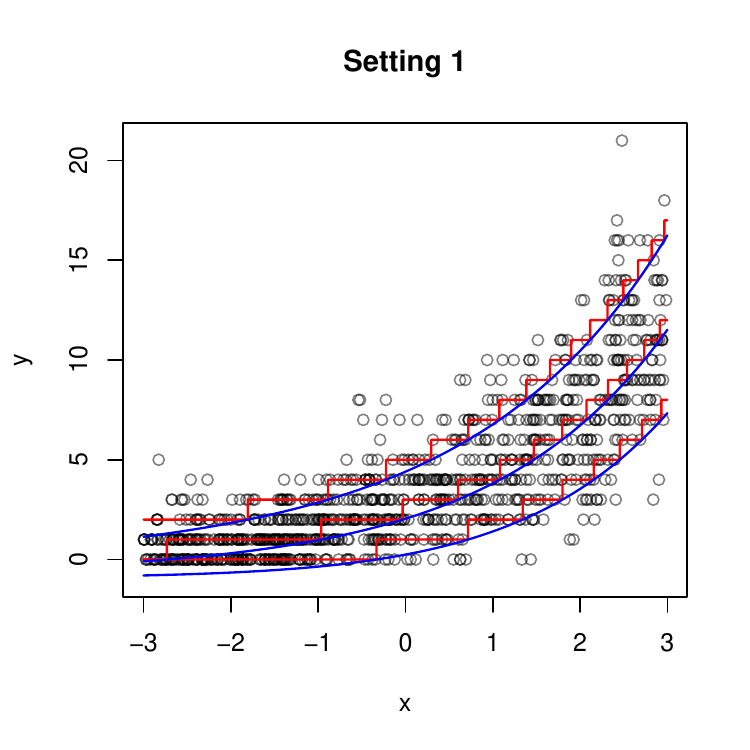}
    \includegraphics[width=0.32\textwidth]{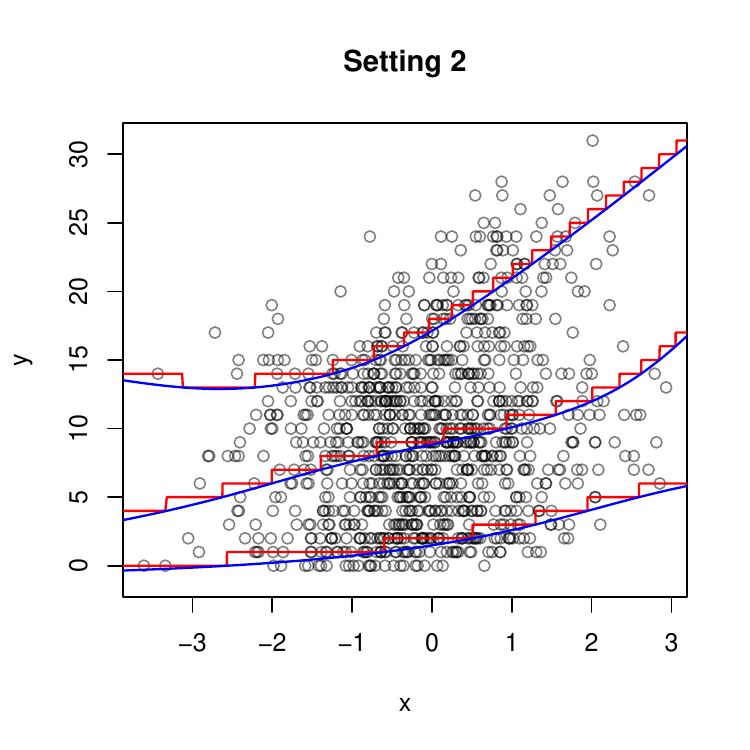}
    \includegraphics[width=0.32\textwidth]{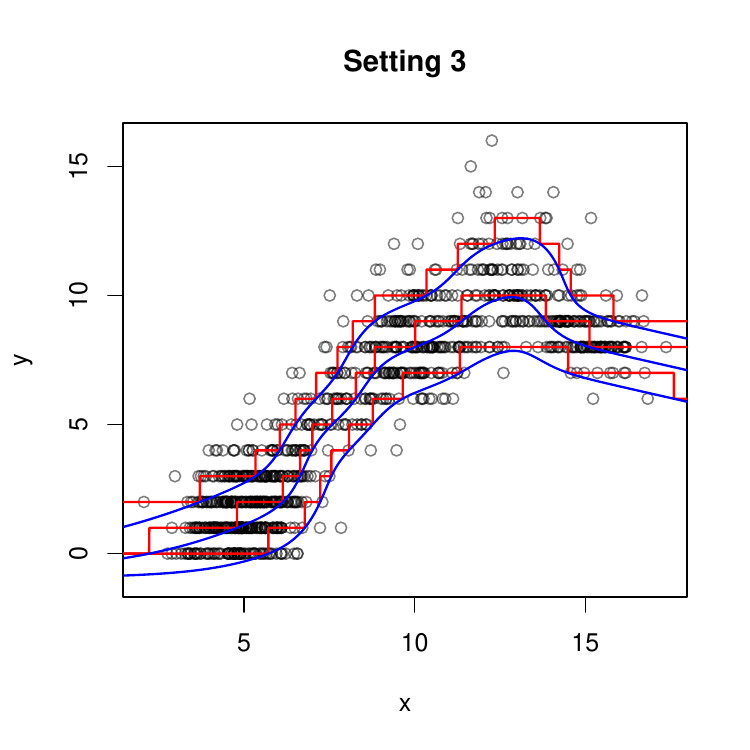}\\
    \includegraphics[width=0.32\textwidth]{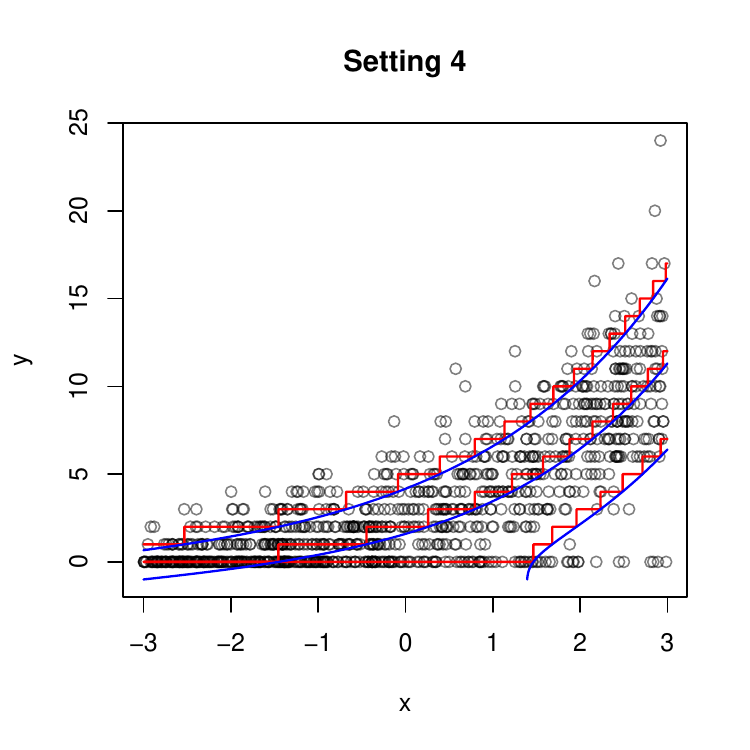}
    \includegraphics[width=0.32\textwidth]{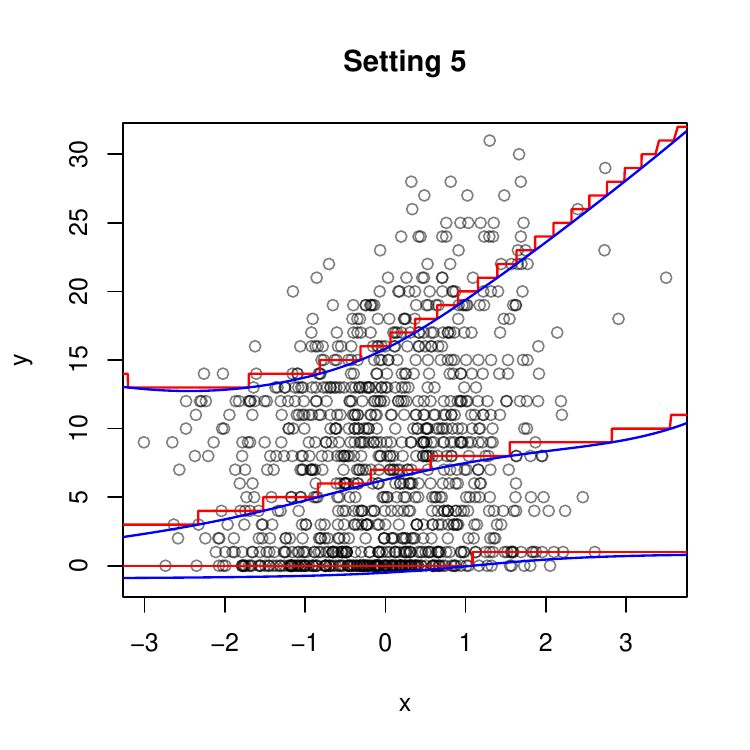}
    \includegraphics[width=0.32\textwidth]{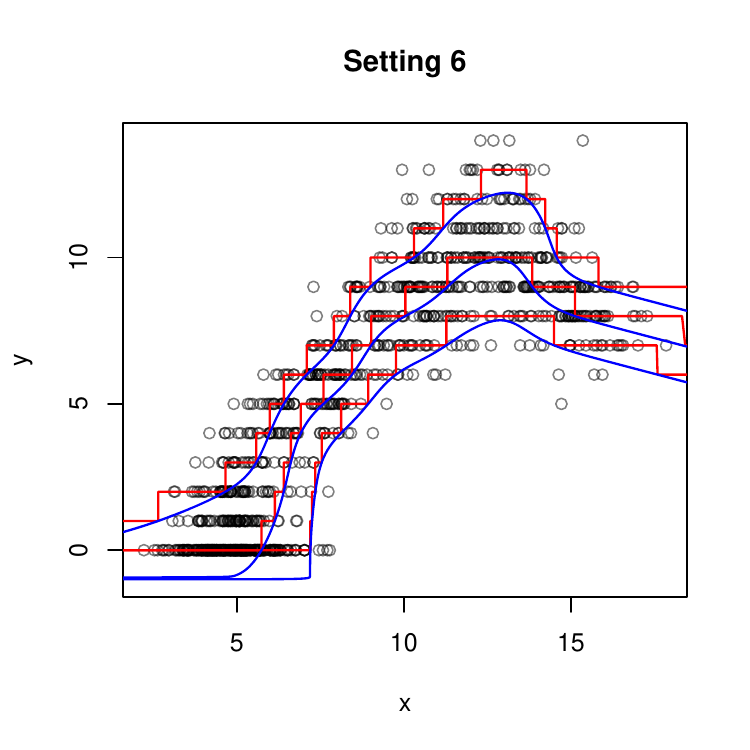}\\
    \includegraphics[width=0.32\textwidth]{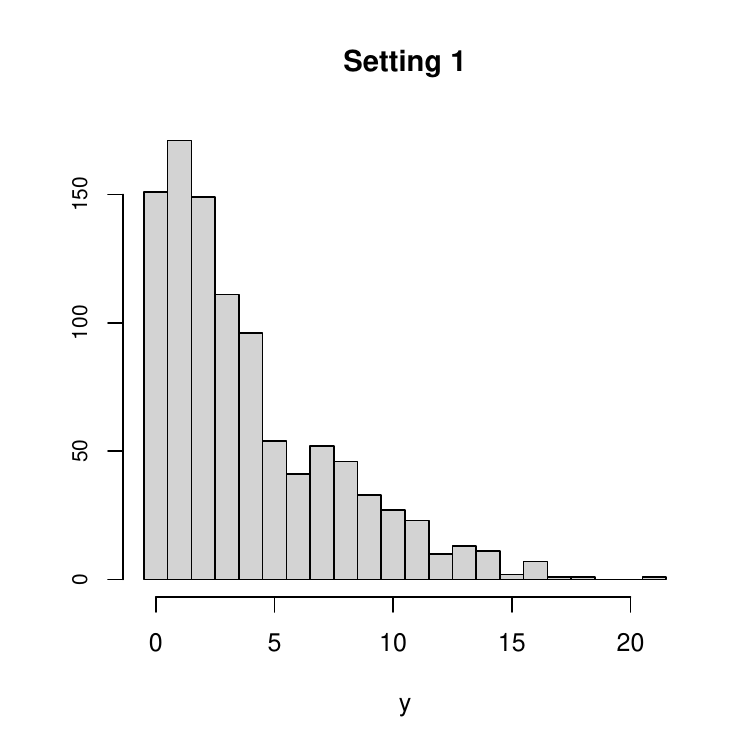}
    \includegraphics[width=0.32\textwidth]{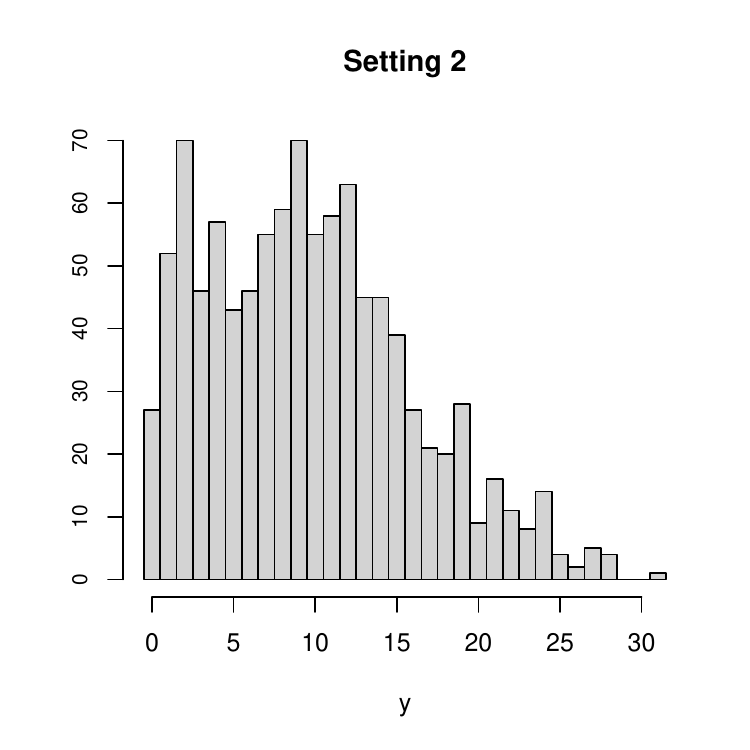}
    \includegraphics[width=0.32\textwidth]{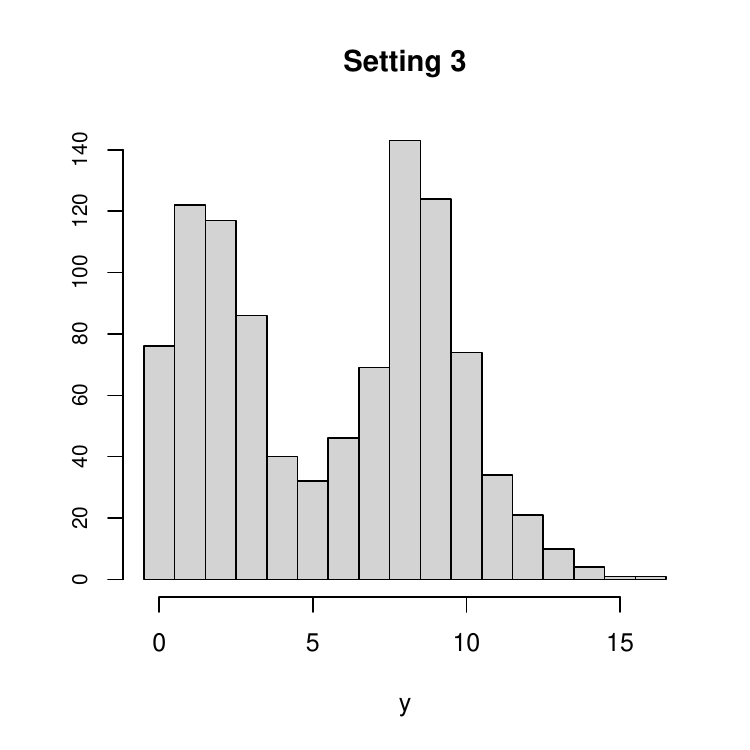}\\
    \includegraphics[width=0.32\textwidth]{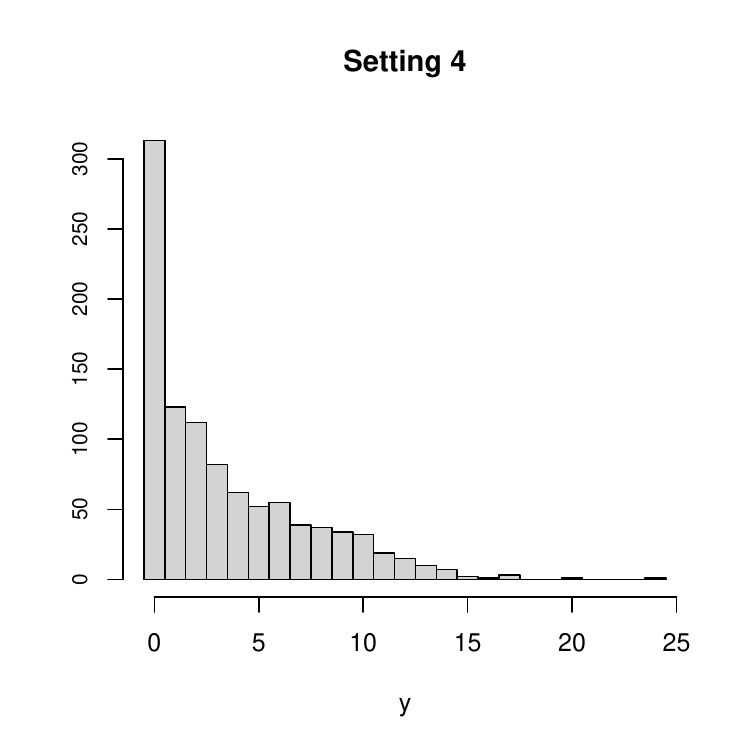}
    \includegraphics[width=0.32\textwidth]{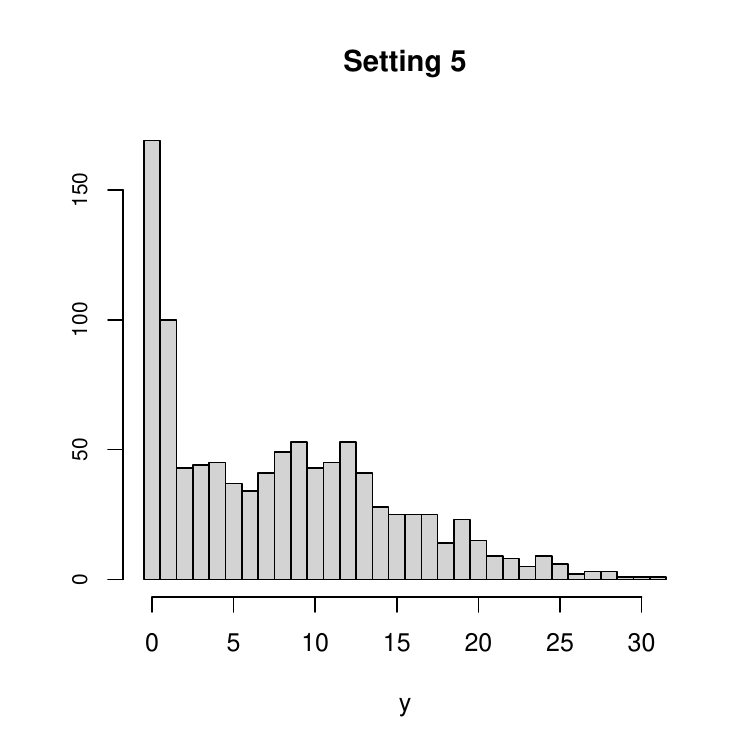}
    \includegraphics[width=0.32\textwidth]{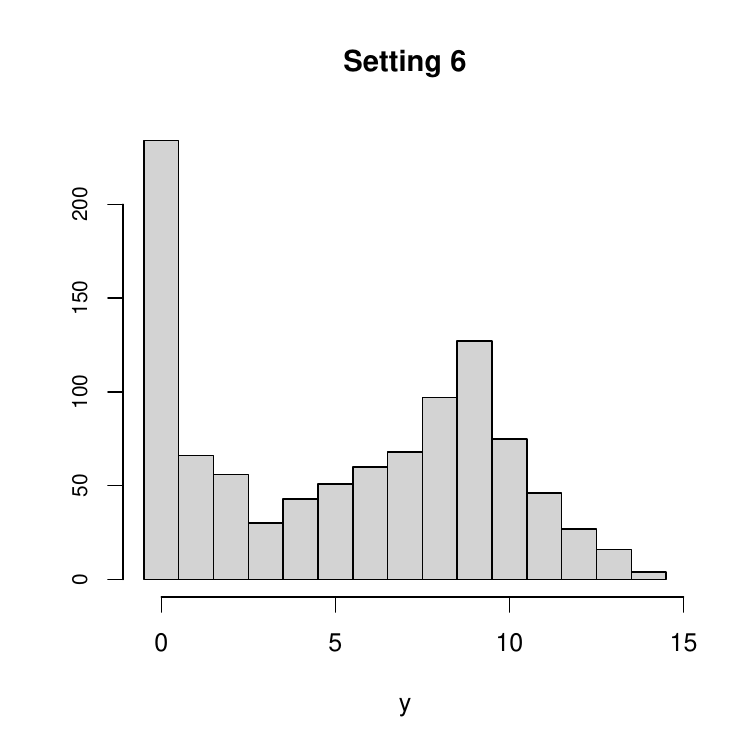}
    \caption{Scatter plots and histograms for the simulated data. 
    The blue lines represent the true conditional continuous quantiles $y_\tau^{*\text{(true)}}$ and red lines represent the count quantiles $Q^\text{(true)}_y(\tau;\vx)=\lceil y_\tau^{*\text{(true)}}\rceil$ for $\tau=0.1$, $0.5$, and $0.9$.}
    \label{fig:simdata}
\end{figure}

\begin{figure}[H]
    \centering
    \includegraphics[width=0.32\textwidth]{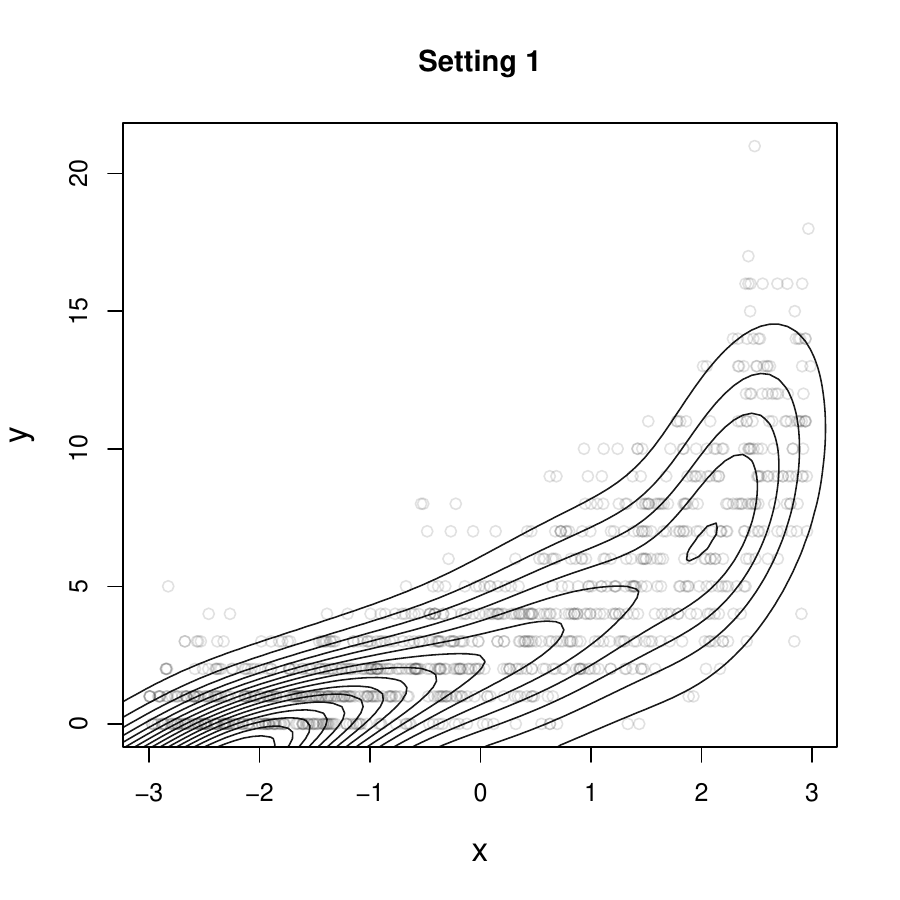}
    \includegraphics[width=0.32\textwidth]{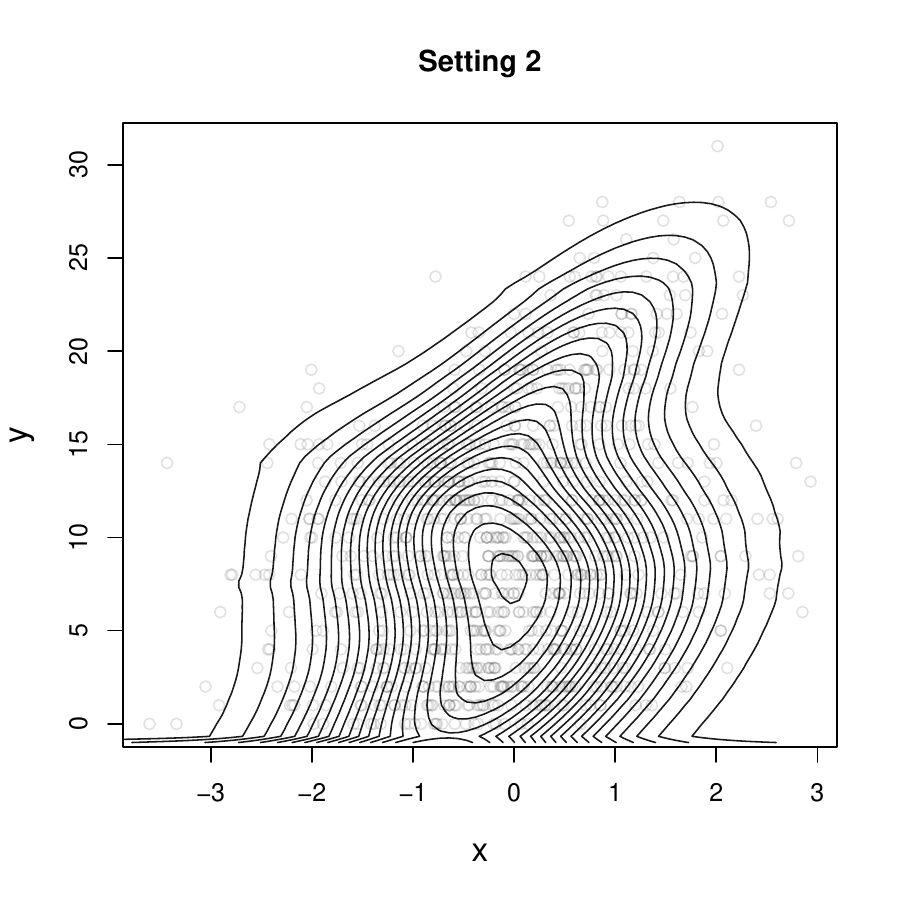}
    \includegraphics[width=0.32\textwidth]{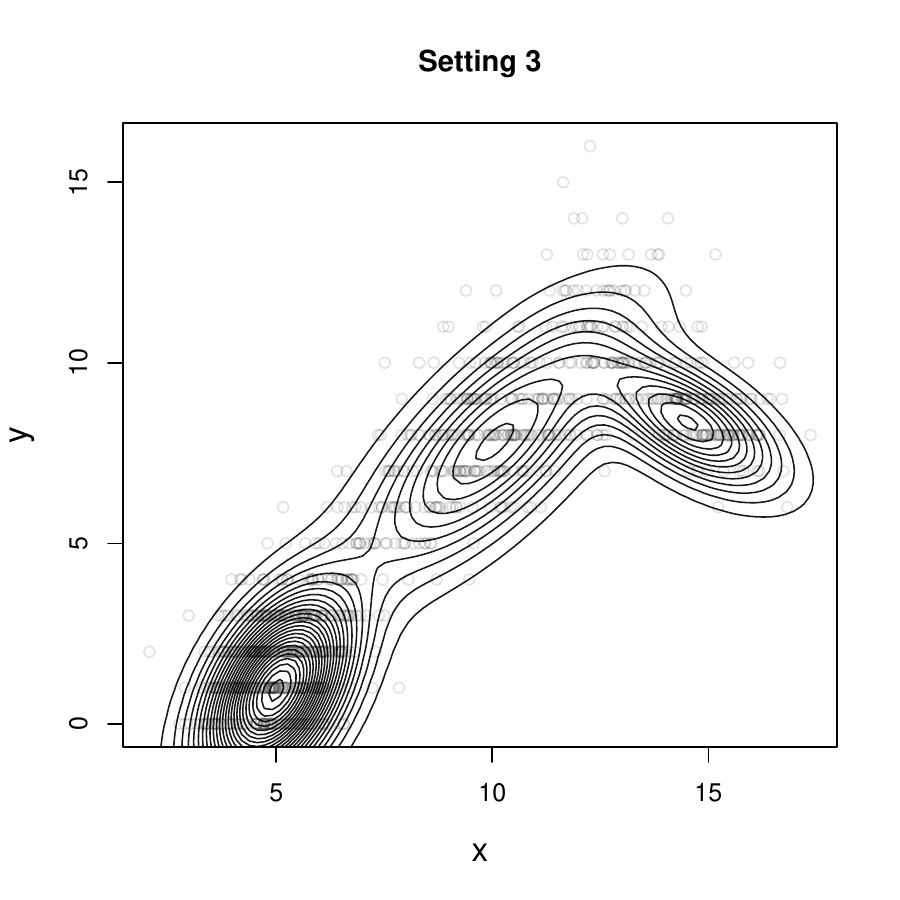}\\
    \includegraphics[width=0.32\textwidth]{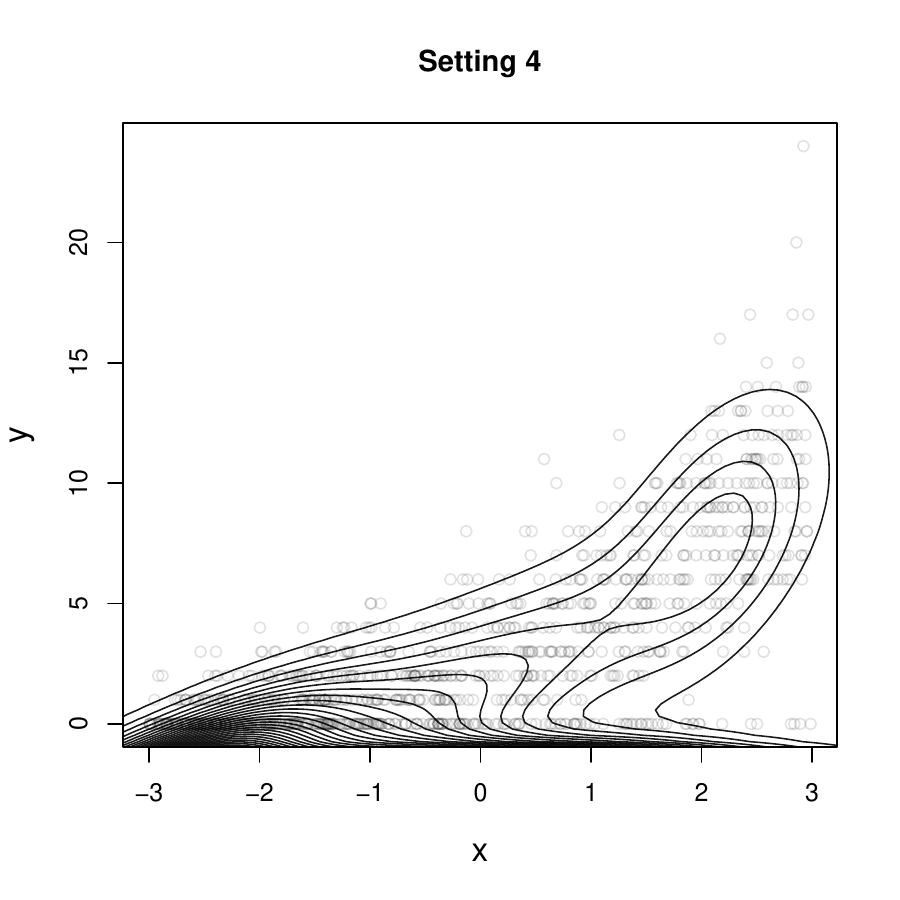}
    \includegraphics[width=0.32\textwidth]{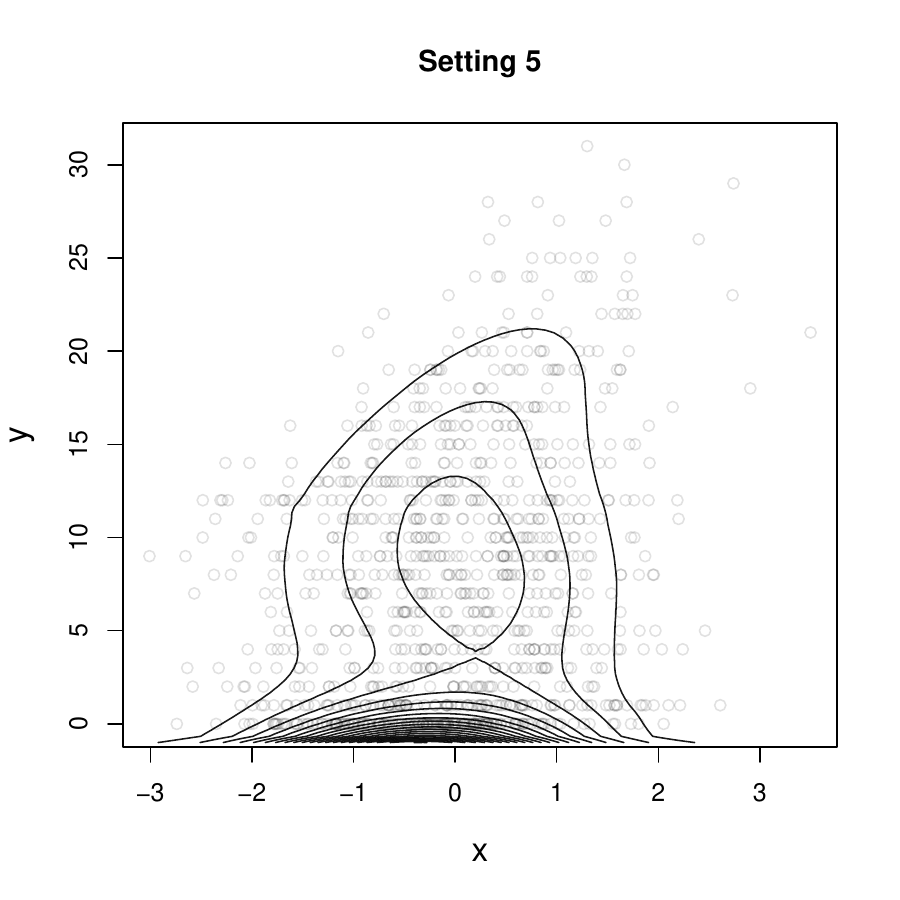}
    \includegraphics[width=0.32\textwidth]{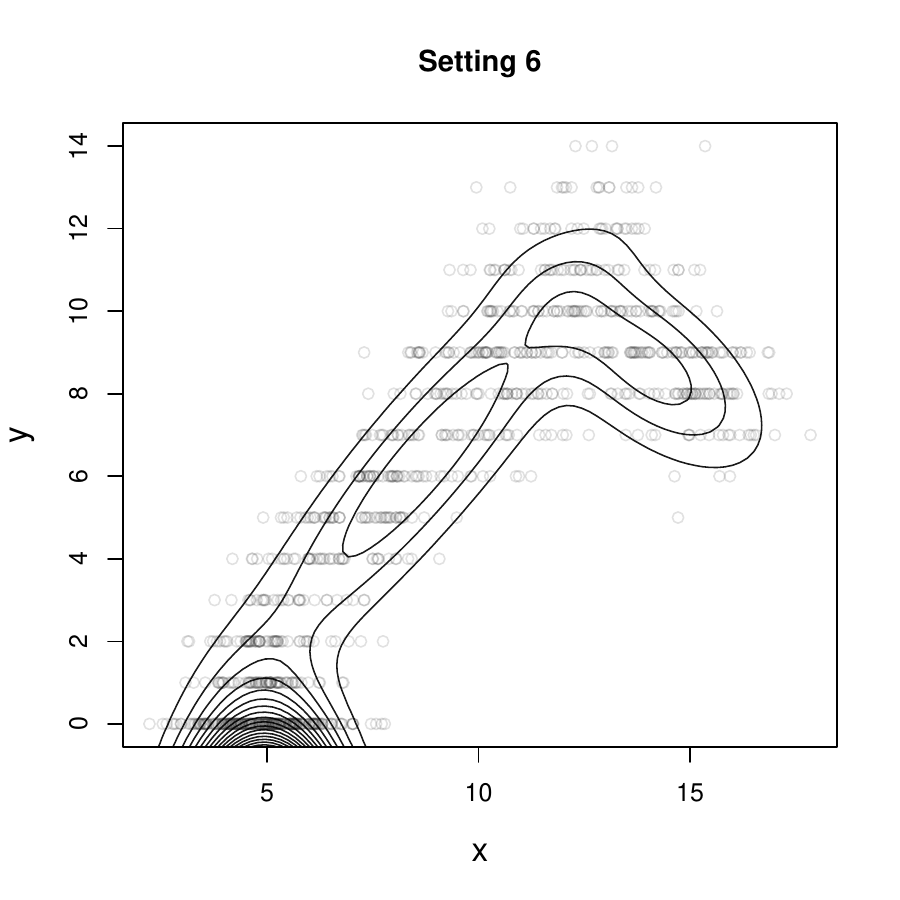}    \caption{Posterior means of the joint densities under the proposed nonparametric model.}
    \label{fig:post_density}
\end{figure}

\begin{figure}[H]
    \centering
    \includegraphics[width=0.32\linewidth]{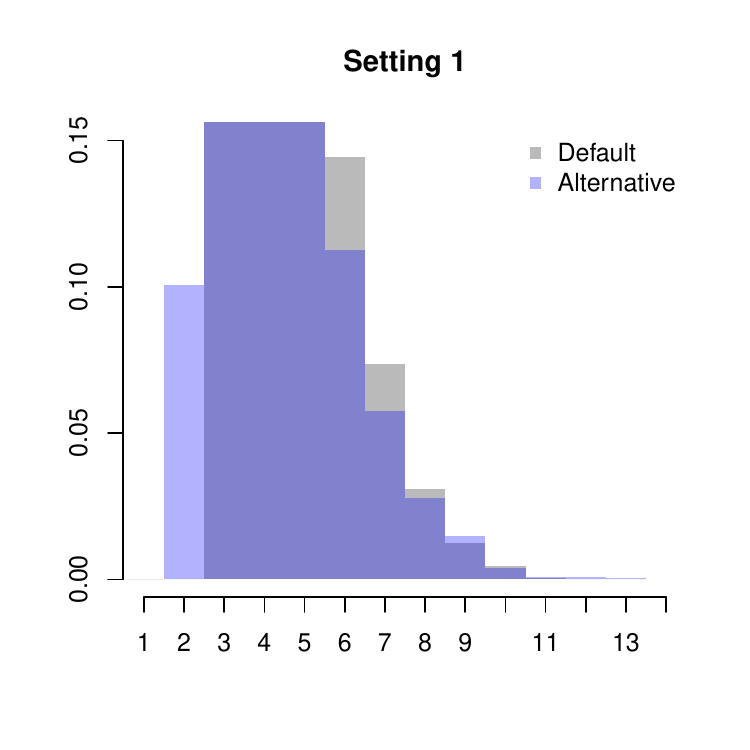}
    \includegraphics[width=0.32\linewidth]{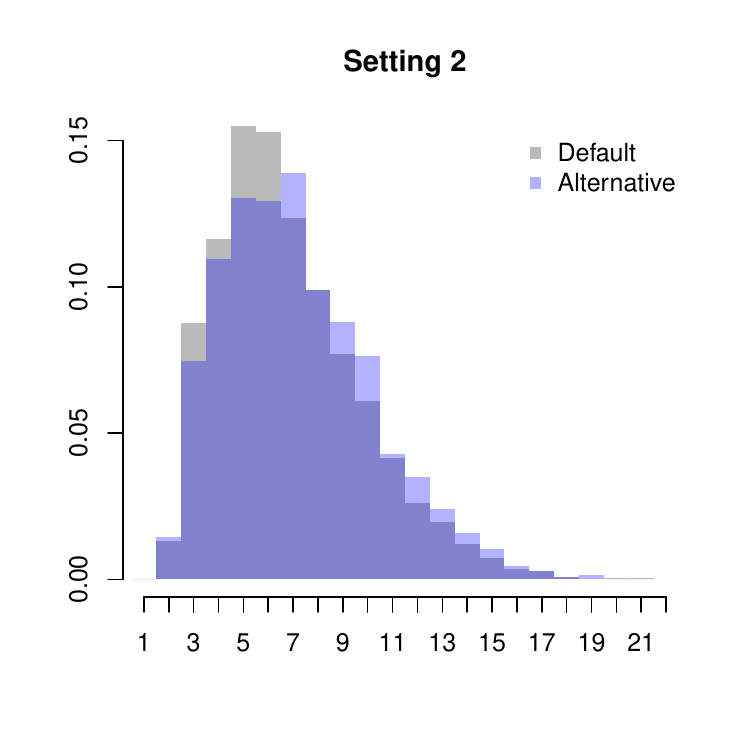}
    \includegraphics[width=0.32\linewidth]{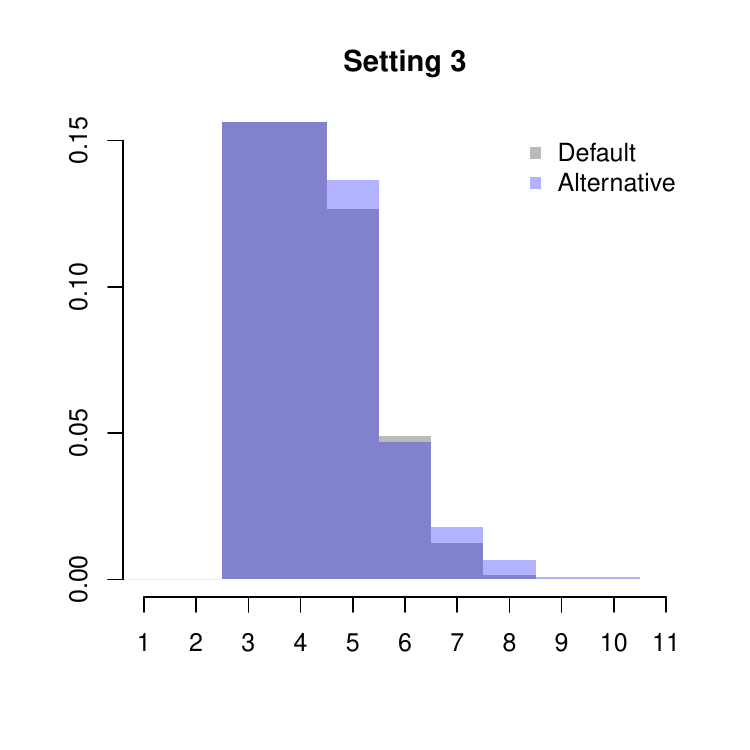}\\
    \includegraphics[width=0.33\linewidth]{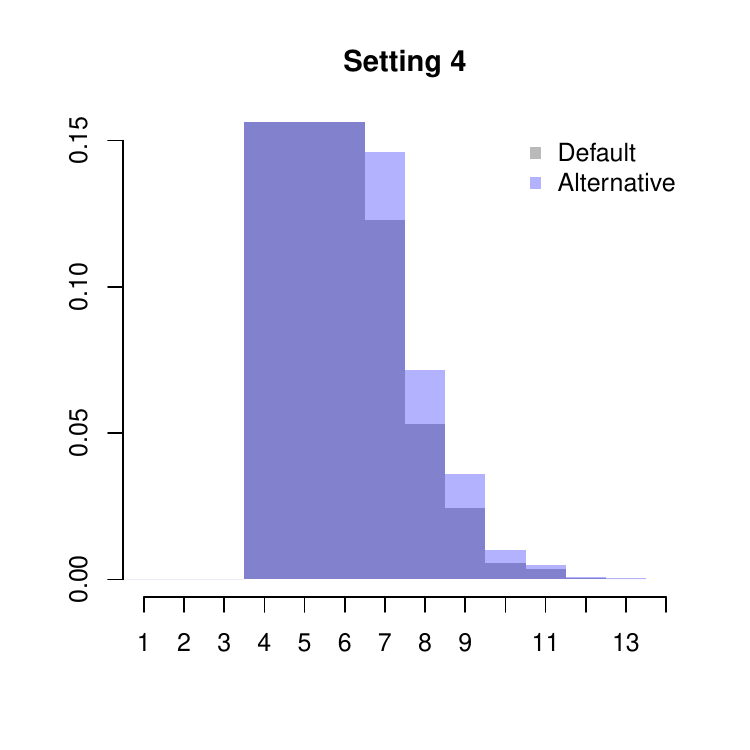}
    \includegraphics[width=0.32\linewidth]{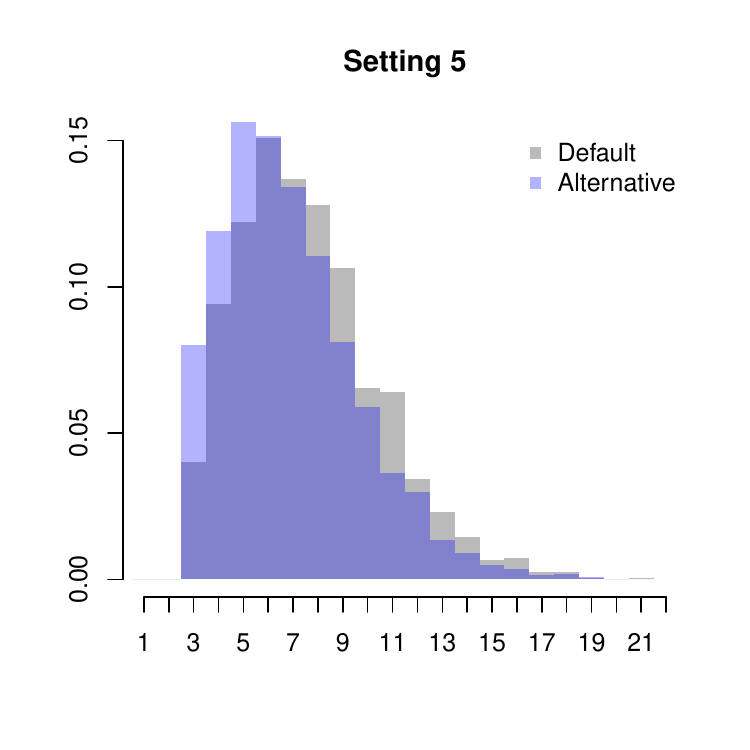}
    \includegraphics[width=0.32\linewidth]{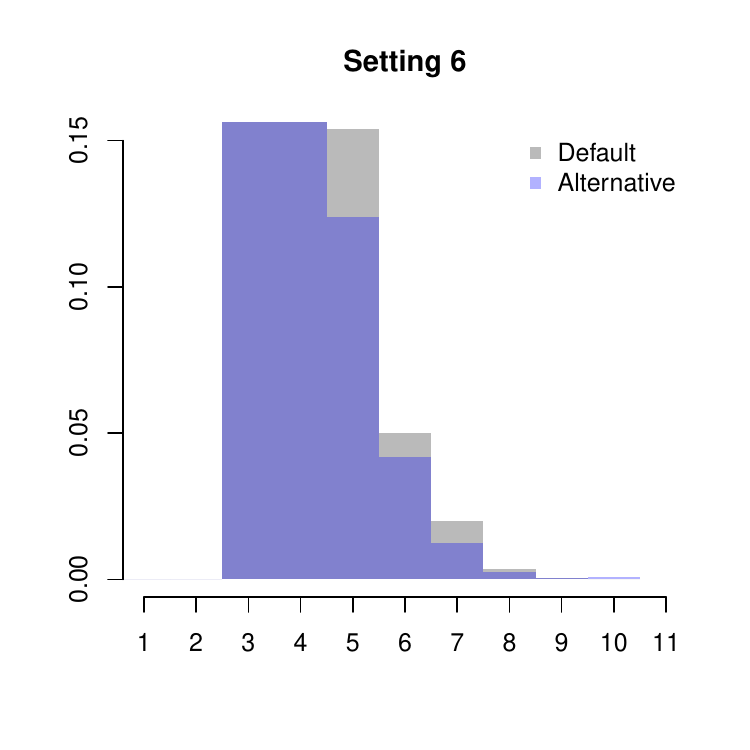}
    \caption{Posterior distributions of numbers of clusters for $\tau=0.1, 0.5$ and $0.9$ under the default and alternative prior specifications for the PY process.}
    \label{fig:alt_k}
\end{figure}

\begin{figure}[H]
    \centering
    \includegraphics[width=0.32\linewidth]{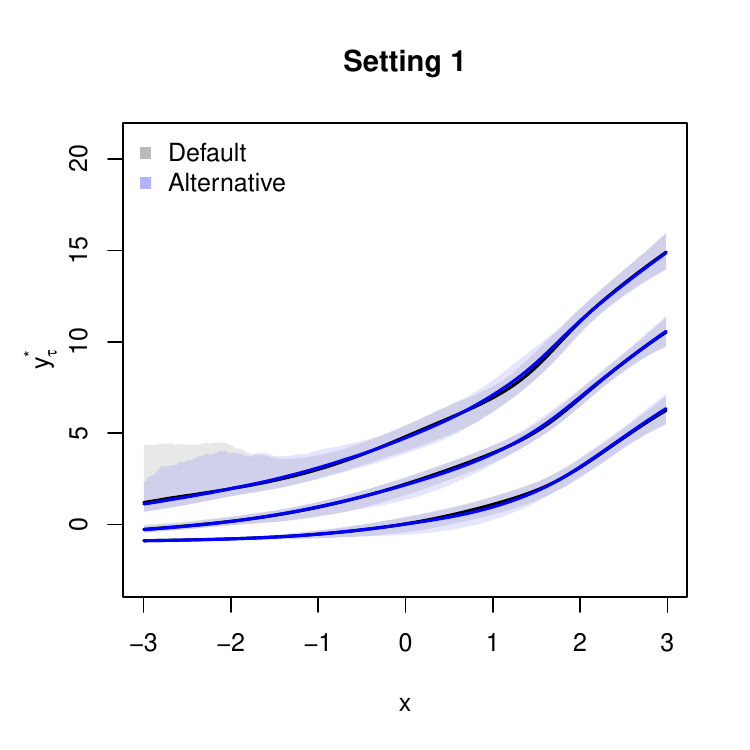}
    \includegraphics[width=0.32\linewidth]{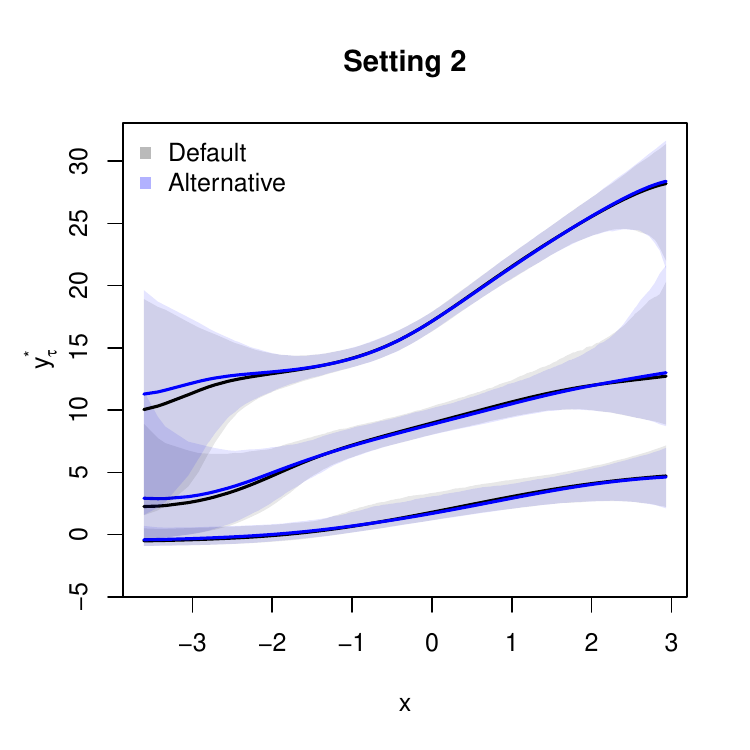}
    \includegraphics[width=0.32\linewidth]{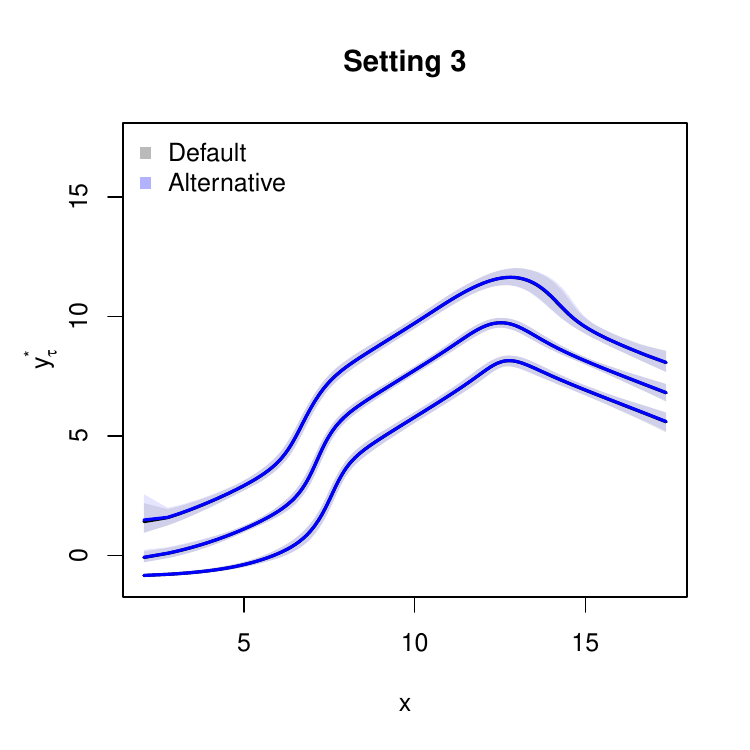}\\
    \includegraphics[width=0.32\linewidth]{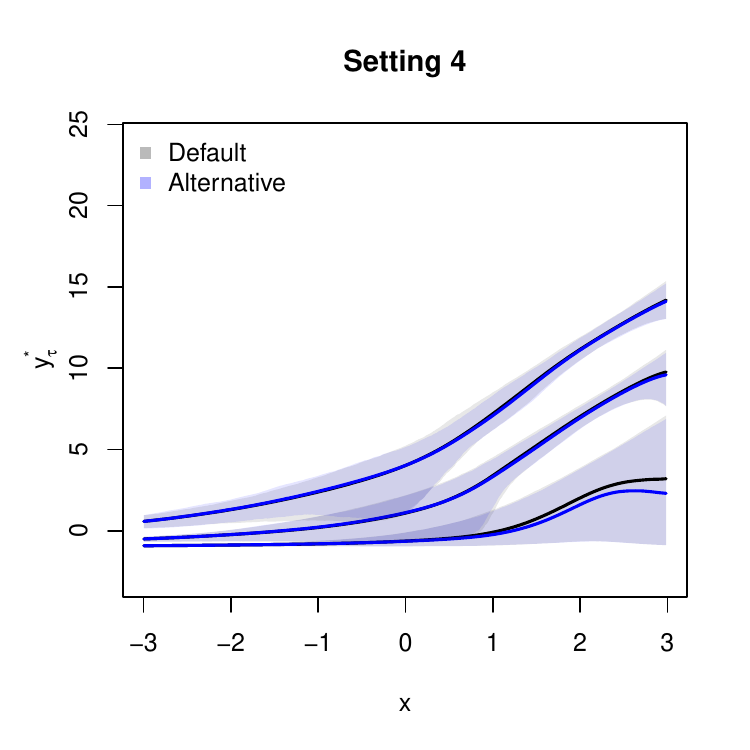}
    \includegraphics[width=0.32\linewidth]{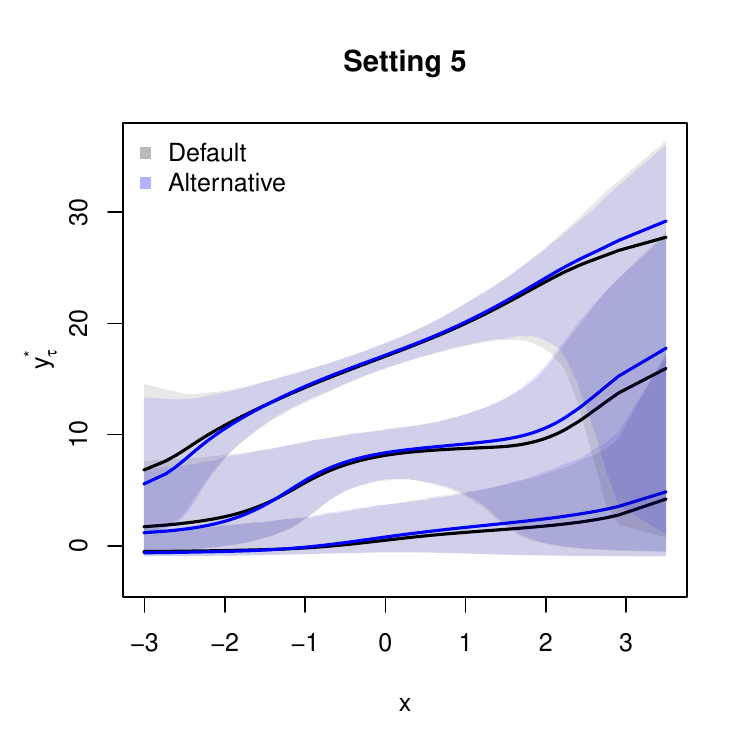}
    \includegraphics[width=0.32\linewidth]{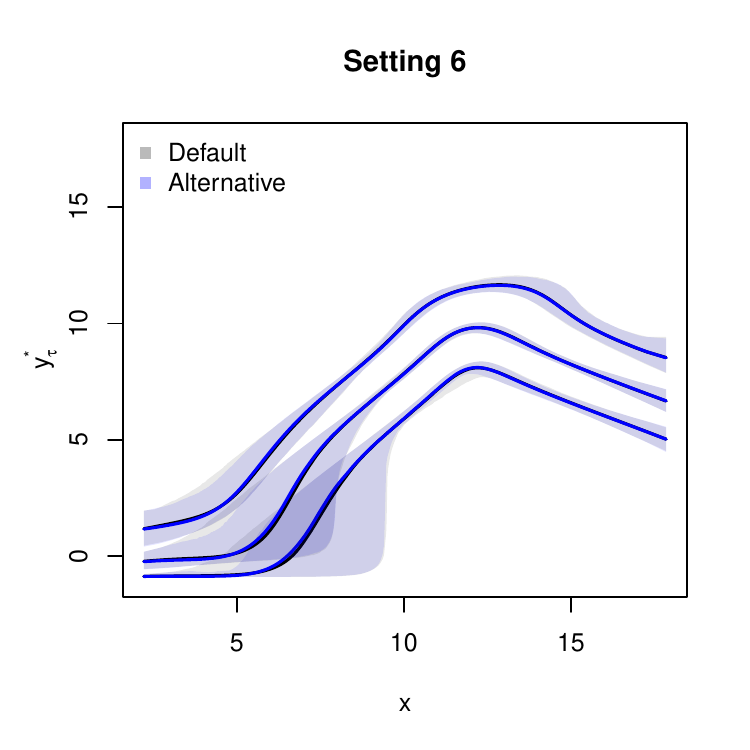}
    \caption{Posterior distributions of $y^*_\tau$ for $\tau=0.1, 0.5$ and $0.9$ under the default and alternative prior specifications for the PY process.}
    \label{fig:alt_reg}
\end{figure}

\begin{figure}[H]
    \centering
    \includegraphics[width=\textwidth]{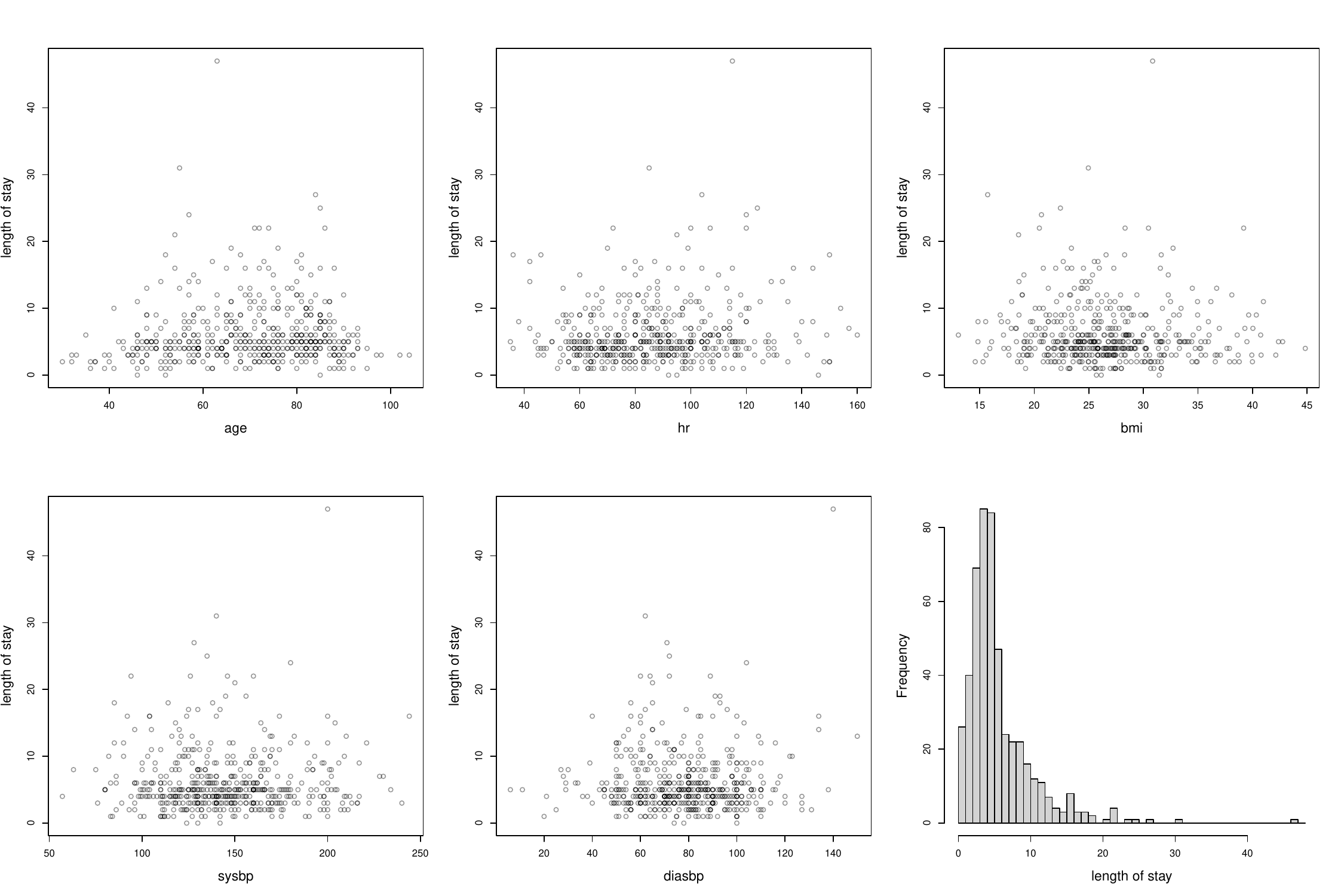}
    \caption{Scatter plots and histogram for WHAS data.}
    \label{fig:whas}
\end{figure}

\begin{table}[H]
    \centering
    \caption{Summary statistics of WHAS data.}
    \begin{tabular}{crrrrrr}\toprule
         & \age & \hr & \bmi & \sysbp & \diasbp & \los \\\hline
         Mean& 69.821 & 86.865 & 26.612 & 144.578 &  78.014 &   6.112 \\
         SD &14.513 & 23.189 &  5.415 & 32.177 & 20.906 &  4.721\\
         \bottomrule
    \end{tabular}
    \label{tab:summary}
\end{table}

\begin{figure}
    \centering
    \includegraphics[height=0.95\textheight]{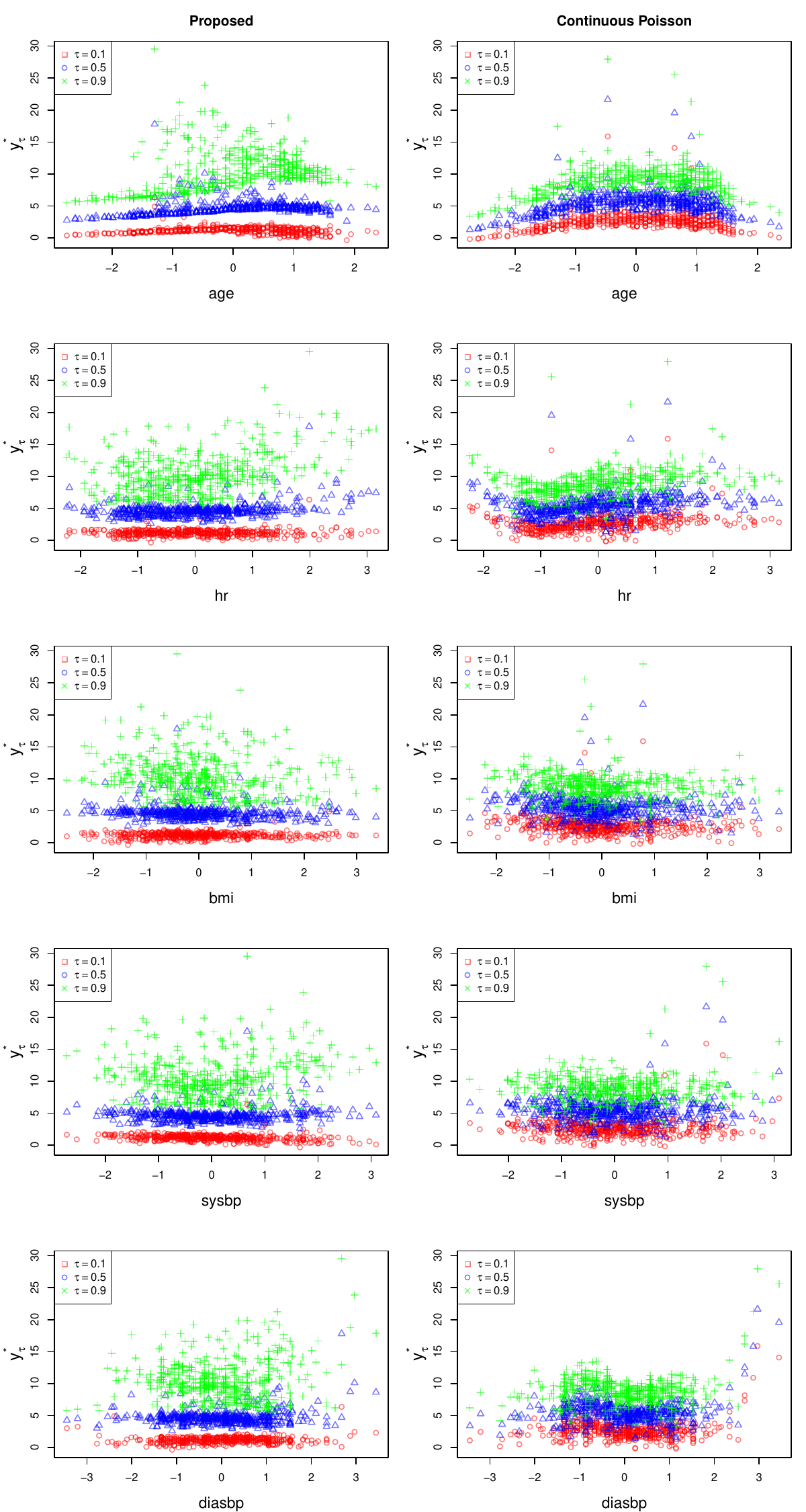}
    \caption{Scatter plots of the posterior means of the conditional continuous quantiles $y^*_\tau$ and the observed covariates $x_j$ for the proposed (left) and continuous Poisson (right) approaches.}
    \label{fig:sp-ytau}
\end{figure}

\begin{figure}[H]
    \centering
    \includegraphics[width=\textwidth]{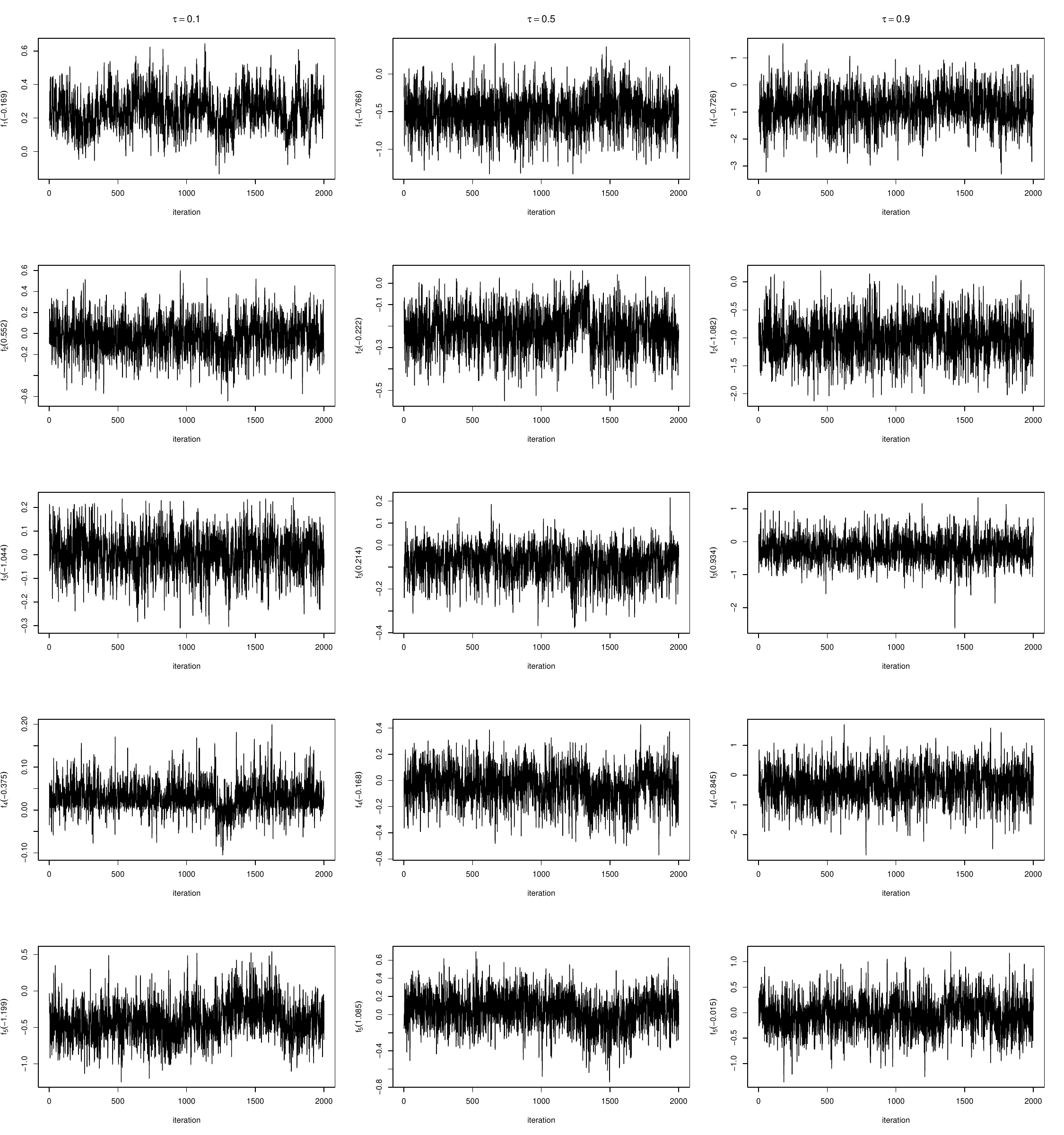}
    \caption{Trace plots of the covrariate effects, $f_{\tau,j}(x_j)=\mB(x_{j})'\vbeta_{\tau,j}$,  at arbitrary selected points for $\tau=0.1$, $0.5$, and $0.9$.}
    \label{fig:trace}
\end{figure}

\end{appendices}

\bibliographystyle{chicago}
\bibliography{ref_count_quantile}

\end{document}